\documentclass[aps, prd, twocolumn, showpacs, showkeys, superscriptaddress, preprintnumbers, floatfix]{revtex4}
\usepackage{multirow}
\usepackage{slashed}
\usepackage{graphicx}
\usepackage{amsmath}
\usepackage{bm}
\usepackage{yhmath}
\usepackage{hyperref}
\usepackage{subfigure}
\usepackage[T1]{fontenc}
\usepackage{color}
\usepackage{cases}
\usepackage{subfigure}
\usepackage{dcolumn, booktabs, bm}
\usepackage{amsfonts, amssymb, stmaryrd, latexsym, amsmath}
\usepackage{textcomp}
\usepackage{amsthm}
\usepackage{mathrsfs}

\usepackage{array}

\newcounter{RomanNumber}

\newcommand{\lyxmathsym}[1]{\ifmmode\begingroup\def\b@ld{bold}
	\text{\ifx\math@version\b@ld\bfseries\fi#1}\endgroup\else#1\fi}

\renewcommand{\arraystretch}{1.9}
\usepackage{braket}
\allowdisplaybreaks
\begin{document}
	\title{Magnetic moments and axial charges of the octet hidden-charm molecular pentaquark family }
	
	\author{Hao-Song Li}\email{haosongli@nwu.edu.cn}\affiliation{School of Physics, Northwest University, Xian 710127, China}\affiliation{Institute of Modern Physics, Northwest University, Xian 710127, China}\affiliation{Shaanxi Key Laboratory for Theoretical Physics Frontiers, Xian 710127, China}\affiliation{Peng Huanwu Center for Fundamental Theory, Xian 710127, China}	
	\author{Fei Guo}\email{guofei@stumail.nwu.edu.cn}\affiliation{School of Physics, Northwest University, Xian 710127, China}
	\author{Ya-Ding Lei}\email{yadinglei@stumail.nwu.edu.cn}\affiliation{School of Physics, Northwest University, Xian 710127, China}
	\author{Feng Gao}\email{fenggao@stumail.nwu.edu.cn}\affiliation{School of Physics, Northwest University, Xian 710127, China}

	\begin{abstract}
			
	In this work, we calculate the magnetic moment and axial charge of the octet hidden-charm molecular pentaquark
	family in quark model. The Coleman-Glashow sum-rule for the magnetic moments of pentaquark
	family is always fulfilled independently of SU(3) symmetry breaking. In the $ 8_{2f} $ flavor representation, the magnetic moments of hidden-charm molecular pentaquark states with spin configuration $J^{P}=\frac{1}{2}^{-}(\frac{1}{2}^{+}\otimes0^{-})$ are all equal to $\mu_{c}=0.38\mu_{N}$. The axial charges of octet hidden-charm molecular pentaquark states are quite small compared to the axial charge of nucleon. The axial charges of the pentaquark states with the spin configuration $J^{P}=\frac{1}{2}^{-}(\frac{1}{2}^{+}\otimes0^{-})$ in $8_{2f}$ flavor representation are all zero.

	\end{abstract}

	\maketitle
	\thispagestyle{empty}
	
	\section{Introduction}
	\label{sec1}
		In order to classify hadrons constantly discovered in experiments, Gell-Mann~\cite{Gell-Mann:1964ewy} proposed a set of correct and effective classification in 1964, called the quark model. The model divides hadrons into singlet, octet and decuplets under $SU(3)$ symmetry. Theorists predicted the existence of multi-quark states after the establishment of the quark model. So far, many multi-quark states including pentaquark and tetraquark states have been observed in different experiments~\cite{Belle:2003nnu,LHCb:2016axx,BESIII:2016adj,LHCb:2018oeg,LHCb:2015yax,LHCb:2019kea,LHCb:2021chn,LHCb:2020jpq,LHCb:2022ogu}. We use the new naming scheme for the representation of pentaquarks in this paper~\cite{Gershon:2022xnn}.
		
		In 2015, the hidden-charm pentaquark state was firstly observed by LHCb collaboration in the $\Lambda^{0}_b\to{}J/\psi{}K^{-}p$ decay. Two candidates of the hidden-charm pentaquark states were announced  as $P^{N}_{\psi}(4380)^{^{+}}$  and $P^{N}_{\psi}(4450)^{^{+}}$ ~\cite{LHCb:2015yax}. In 2019, three new pentaquark states $P^{N}_{\psi}(4312)^{^{+}}$, $P^{N}_{\psi}(4440)^{^{+}}$  and $P^{N}_{\psi}(4457)^{^{+}}$, were reported in the updated analyses of the LHCb collaboration~\cite{LHCb:2019kea}. The $P^{N}_{\psi}(4450)^{^{+}}  $  reported earlier is split into two peaks corresponding to the $P^{N}_{\psi}(4440)^{^{+}}  $  and the $P^{N}_{\psi}(4457)^{^{+}}$. In 2020, the LHCb Collaboration observed the hidden-charm strange pentaquark state $P_{\psi{}s}^{\Lambda}(4459)^{0}$ in the $J/\psi{}\Lambda$ mass spectrum through an amplitude analysis of the $\Xi^{-}_b\to{}J/\psi{}\Lambda{}K^{-}$ decay \cite{LHCb:2020jpq}. In 2021, a new pentaquark structure $P^{N}_{\psi}(4337)^{^{+}}  $  was observed by the LHCb Collaboration in the $B_{s}^{0}\to J/\psi p\bar{p}$ decay~\cite{LHCb:2021chn}. Recently, the LHCb collaboration observed a new structure $P_{\psi{}s}^{\Lambda}(4338)^{0}$ in the $B^{-}\to{}J/\psi{}\Lambda\bar{p}$ decay~\cite{LHCb:2022ogu}.
	
		With the discovery of the hidden-charm pentaquark states in the experiments, many theorists have predicted the spin-parity quantum numbers of the hidden-charm pentaquark states~\cite{Deng:2022vkv,Wang:2019nvm,Xiao:2019gjd,Liu:2020hcv,Peng:2020hql,Zhu:2021lhd,Du:2021bgb,Chen:2021tip,Yang:2021pio,Ozdem:2021btf},  although the reliable information about their spin-parity has been unavailable until now. 
		The near-threshold behaviors of these states in Table \ref{tab_1} indicate that those candidates of the hidden-charm pentaquark states can be explained as the molecular state belonging to the octet in the quark model. It is reasonable to speculate that the remaining hidden-charm pentaquark states will be observed in the coming future.

	\renewcommand{\arraystretch}{2.0}	
		\begin{table}[htbp]
			\centering
			\caption{Properties of the hidden-charm pentaquark states}
			\vspace{0.5em}
			\label{tab_1} 
			\resizebox{86mm}{!}{
			\begin{tabular}{ c|c|c|c}
			\toprule[1.0pt]
			\toprule[1.0pt]
				Pentaquarks & Masses$(\mbox{MeV})$ & Widths$(\mbox{MeV})$  & Near Thresholds
				\\
				\hline
				$P^{N}_{\psi}(4380)^{^{+}}$ & $4380\pm8\pm29$ &$205\pm18\pm86$ &  $\Sigma_{c}^{*}\bar{D}$
				\\
				\hline
				$P^{N}_{\psi}(4312)^{^{+}}$ & $4311.9\pm0.7\ ^{+6.8}_{-0.6}$ &$9.8\pm2.7\ ^{+3.7}_{-4.5}$ &  $\Sigma_{c}\bar{D}$
				\\
				\hline
					$P^{N}_{\psi}(4440)^{^{+}}$ & $4440.3\pm1.3\ ^{+4.1}_{-4.7}$ &$20.6\pm4.9\ ^{+8.7}_{-10.1}$ &   $\Sigma_{c}\bar{D}^{*}$
				\\
				\hline
					$P^{N}_{\psi}(4457)^{^{+}}$ & $4457.3\pm0.6\ ^{+4.1}_{-1.7}$ &$6.4\pm2.0\ ^{+5.7}_{-1.9}$ &  $\Sigma_{c}\bar{D}^{*}$
				\\
				\hline
				$P^{N}_{\psi}(4337)^{^{+}}$ & $4337.3\ ^{+7}_{-4}\ ^{+2}_{-2}$ &$29\ ^{+26}_{-12}\ ^{+14}_{-14}$ &  $\Lambda_{c}\bar{D}^{*}$
				\\
				\hline
			$P_{\psi{}s}^{\Lambda}(4459)^{0}$ & $4458.8\pm2.9\ ^{+4.7}_{-1.1}$ &$17.3\pm6.5\ ^{+8.0}_{-5.7}$ &  $\Xi_{c}^{}D^{*}$
				\\
				\hline
			$P_{\psi{}s}^{\Lambda}(4338)^{0}$ & $4338.2\pm0.7\pm0.4$ &$7.0\pm1.2\pm1.3$ &  $\Xi_{c}^{}D^{}$	
				\\
				\toprule[1.0pt]
				\toprule[1.0pt]
			\end{tabular}}
		\end{table}	
	
		In hadron physics, it is interesting and important to investigate the properties of the hidden-charm pentaquark family in molecular models, which provides useful clues about the internal structure of the pentaquark states. The electromagnetic properties of hadrons are crucial for understanding their strong interactions and structure, which help to gain insight into QCD in the low-energy regime. The magnetic moment of hadrons encodes important details about the internal charge and magnetization distributions. In Ref.~\cite{Gao:2021hmv,Wang:2016dzu,Guo:2023fih}, the authors calculated the magnetic moments of hidden-charm pentaquark states with  $J^P={\frac{1}{2}}^{\pm}$, ${\frac{3}{2}}^{\pm}$, ${\frac{5}{2}}^{\pm}$, and ${\frac{7}{2}}^{+}$ in quark model.  In Ref.~\cite{Ozdem:2018qeh,Ozdem:2021btf,Ozdem:2021ugy}, within the framework of QCD light-cone sum rules, the authors extracted the magnetic dipole moment of the 	$P^{N}_{\psi}(4312)^{^{+}}$,	$P^{N}_{\psi}(4440)^{^{+}}$, 	$P^{N}_{\psi}(4457)^{^{+}}$, 	$P^{N}_{\psi}(4380)^{^{+}}$ and 	$P_{\psi{}s}^{\Lambda}(4459)^{0}$ pentaquark states in the molecular and diquark-diquark-antiquark models. In Ref.~\cite{Ortiz-Pacheco:2018ccl}, the authors obtained the ground state of hidden-charm pentaquark states with $J^P={\frac{3}{2}}^{-}$ quantum numbers as well as the related magnetic dipole moment and electromagnetic coupling constant, which is significant for quark-model-driven pentaquark photoproduction experiments.	
	
		The axial charge $g_A$ of baryons is a fundamental quantity for understanding the electroweak and strong interactions in standard model~\cite{UCNA:2012fhw,Mund:2012fq}. It not only controls weak interaction processes, such as $\beta$ decay, but also intertwines weak and strong interactions.  The Goldberger-Treiman relation~\cite{Goldberger:1958tr}, $g_A=f_{\pi}g_{\pi NN}/M_N$, reflects this most clearly. When the $\pi$ decay constant $f_{\pi}$ and nucleon mass $M_N$ are given, the $\pi NN$ coupling constant $g_{\pi NN}$ is proportional to $g_{A}$. Therefore, the correlation between $\pi$ degrees of freedom and axial charges is closely related to hadron physics. $g_A$ also represent an indicator of non-perturbative QCD chiral symmetry breaking, thus, the axial charge is an important parameter in low-energy effective theories. Therefore, investigating the axial charge of the octet hidden-charm pentaquark family can serve as another effective approach independent of the magnetic moment.
	
		Studying the magnetic moment and axial charge of pentaquark hadrons is so important that they provide insight into the internal structure and the interactions between constituent quarks, which help to understand the fundamental strong force that binds quarks together to hadrons. With the continuous advancement of the experimental process in the future,  discoveries of the octet hidden-charm molecular pentaquark
		family are no longer far away. Researchers will be more eager to explore information about  internal structure of the  pentaquark	states, and research on magnetic moment and axial charge can provide a large amount of data reference for this. Therefore, in this paper, we calculate the magnetic moment and axial charge of the octet hidden-charm molecular pentaquark states with quark model. 
	
		The remainder of this paper is organized as follows. In Sec.\ref{sec2} , we introduce the wave function in molecular model, while in Sec.\ref{sec3}, we calculate the magnetic moment of octet hidden-charm molecular pentaquark family. In Sec.\ref{sec4}, we calculate the axial charge of octet hidden-charm molecular pentaquark family. Sec.\ref{sec5} summarizes this work.

	\section{Wave Functions}  
	\label{sec2}
	The wave function $\Psi$ of the hadron state is a prerequisite for exploring its electromagnetic properties. The overall wave function of the hadron state is composed of the flavor wave function $\psi_{flavor}  $, spin wave function $ \chi_{spin} $,  color wave function $ \xi_{color} $ and space wave function $ \eta_{space} $ composition. The specific representation is
	\begin{align}
	\Psi=\psi_{flavor}\otimes\chi_{spin}\otimes\xi_{color}\otimes\eta_{space}.
	\end{align}
 
 The Fermi statistic requires the overall  wave function to be antisymmetric. Due to the color wave function $ \xi_{color} $ is antisymmetric and the space wave function $ \eta_{space} $ is symmetric in the ground state, we need to consider the symmetry requirement of the flavor-spin wave function when calculating the magnetic moment.

 The hidden-charm pentaquark family in the molecular model is composed of the corresponding singly charmed baryons and anti-charmed mesons. Singly charmed baryons can be obtained from light baryons under $SU(3)$ symmetry by substituting a light quark for a charm quark. 	The two light quarks in the singly charmed baryons can be symmetrical or anti-symmetrical. In the former case, the flavor of the singly charmed baryons belongs to $6_{f}$, and it can form $10_{f}$ and $8_{1f}$ with the anti-charmed meson(${3}_{f}$). In the latter case, the flavor of the singly charmed baryons belongs to $\bar{3}_{f}$, which can form $8_{2f}$ and $1_{f}$ with anti-charm meson (${3}_{f}$). We obtain the direct product $3\otimes3\otimes3 = 1\oplus8_1\oplus8_2\oplus10$. The hidden-charm pentaquark states which have been observed so far are likely to belong to the octet states. Therefore, it is very meaningful to study the properties of the octet hidden-charm molecular pentaquark states. These octet hidden-charm molecular pentaquark states have three spin configurations $J^{P}(J_{b} ^{P_{b}}\otimes J_{m} ^{P_{m}})  $: $\frac{1}{2}^{-}(\frac{1}{2}^{+}\otimes0^{-})  $, $\frac{1}{2}^{-}(\frac{1}{2}^{+}\otimes1^{-})  $, and $\frac{3}{2}^{-}(\frac{1}{2}^{+}\otimes1^{-})  $,  where $ J^{P} $ is the total spin of  pentaquark states, and   $J_{b} ^{P_{b}}\otimes J_{m} ^{P_{m}} $ are corresponding to the angular momentum and parity of baryon and meson, respectively.
 
 \begin{figure}[htbp]
 	\centering
 	\includegraphics[width=1.0\linewidth]{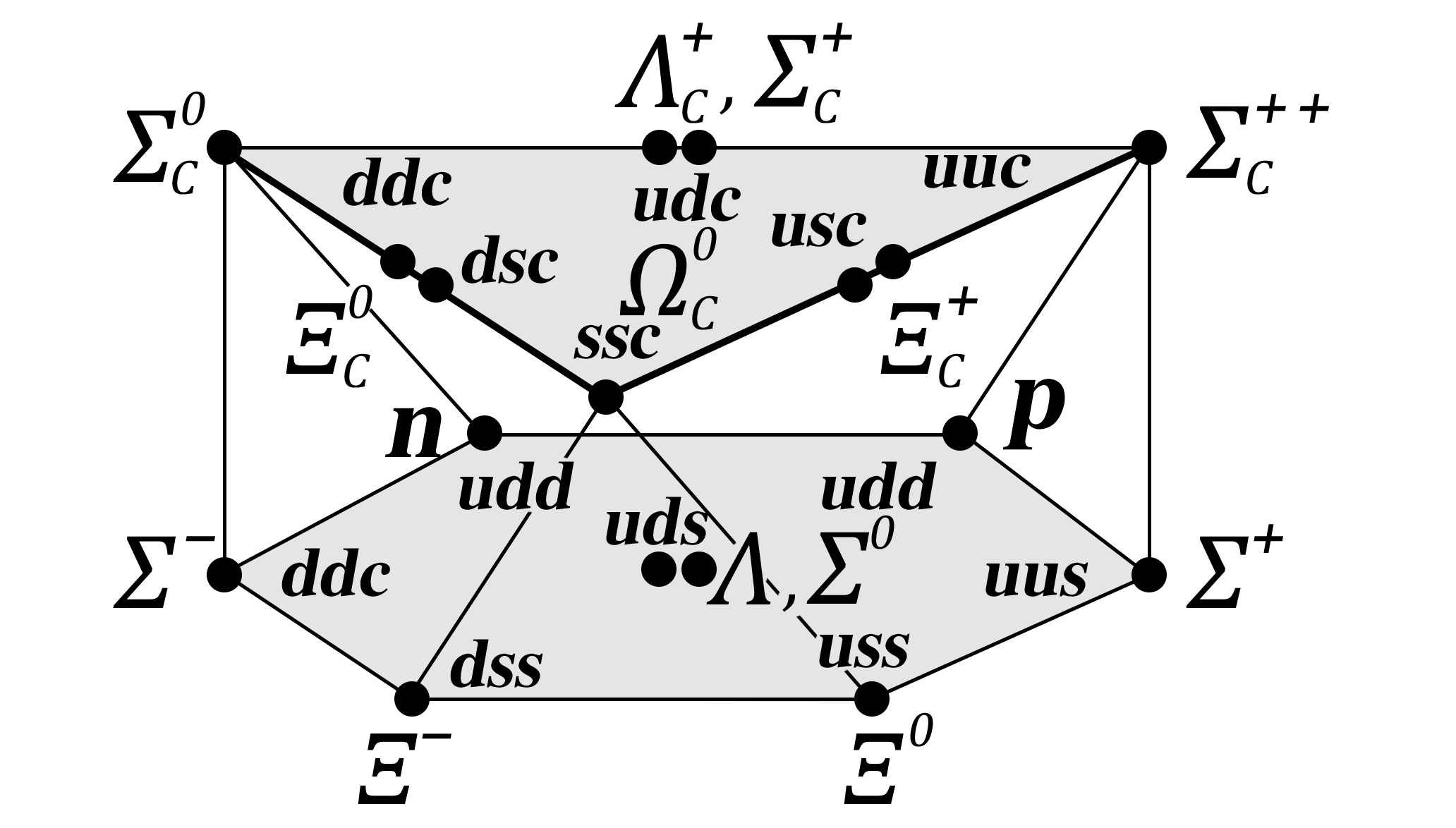}
 	\caption{The octet baryons with $J^{P} = \frac{1}{2}^{+}$ under SU(3) symmetry become singly charmed baryons by replacing a light quark with a charmed quark.}
 	\label{fig_1}
 \end{figure}
 
 \begin{figure}[htbp]
 	\centering
 	\includegraphics[width=0.7\linewidth]{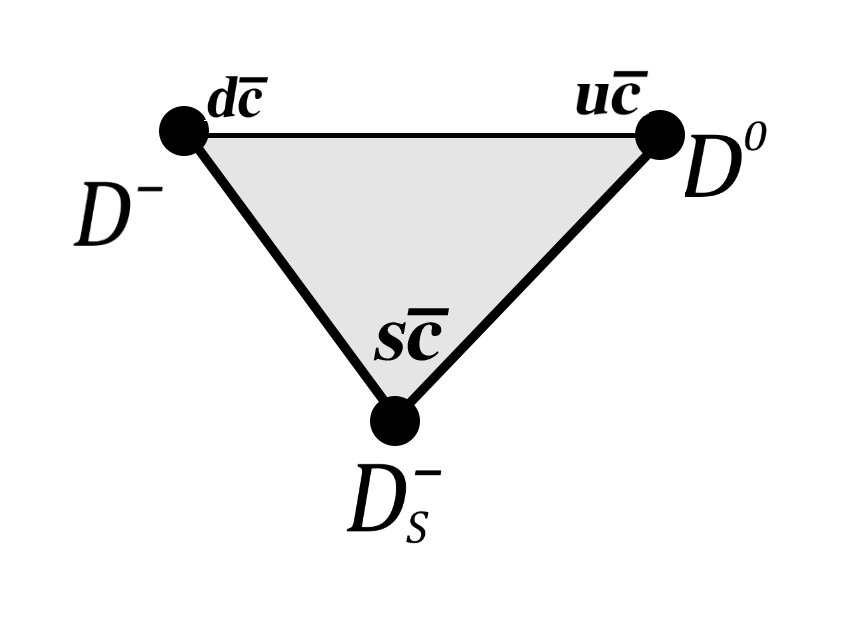}
 	\caption{The anti-charmed mesons are composed of a anti-charm quark and a light quark.}
 	\label{fig_2}
 \end{figure}
 
 In Fig. \ref{fig_1}, we show that the octet baryons with $J^P = \frac{1}{2}^{+}$ become singly charmed baryons by replacing a light quark with a charmed quark. In Fig. \ref{fig_2}, anti-charmed mesons are composed of a anti-charm quark and a light quark.  In Table \ref{tab_3}, we show that the flavor wave functions $\psi_{flavor}$ and the spin wave functions $\chi_{spin}$ of the $S$-wave singly charmed baryons. In Table \ref{tab_4}, we show that the flavor wave functions $\psi_{flavor}$ and the spin wave functions $\chi_{spin}$ of the $ S $-wave anti-charmed mesons. We list the expression for the flavor wave function of the hidden-charm molecular pentaquark under $SU(3)$ symmetry in Table \ref{tab_2}
		\renewcommand{\arraystretch}{1.65}
		\begin{table}[htp]
			\centering
			\caption{The flavor wave functions $\psi_{flavor}$ and  the spin wave functions $\chi_{spin}$ of the $S$-wave singly charmed baryon. Here, $S$ and $S_3$ are the spin and its third component, while	the arrow denotes the third component of the quark spin.}
			\label{tab_3} 
			\begin{tabular}{c|l|l}
			\toprule[1.0pt]
			\toprule[1.0pt]
				Baryons 
				&	$\ket{S,S_3}$  
				&	$\psi_{flavor} \otimes \chi_{spin}$
				\\
				\hline
		
				\multirow{2}{*}{$\Sigma_{c}^{++}$}
				&	$\ket{\frac{1}{2},\frac{1}{2}}$  
				&	$uuc \otimes {\frac{1}{\sqrt{6}}}(2\uparrow\uparrow\downarrow-\downarrow\uparrow\uparrow-\uparrow\downarrow\uparrow)$
				\\
				
				&	$\ket{\frac{1}{2},-\frac{1}{2}}$ 
				&   $uuc \otimes {\frac{1}{\sqrt{6}}}(\uparrow\downarrow\downarrow+\downarrow\uparrow\downarrow-2\downarrow\downarrow\uparrow)$
				\\
				\hline
		
				\multirow{2}{*}{$\Sigma_{c}^{+}$ }
				&	$\ket{\frac{1}{2},\frac{1}{2}}$  
				&   ${\frac{1}{\sqrt{2}}}(udc + duc) \otimes {\frac{1}{\sqrt{6}}}(2\uparrow\uparrow\downarrow-\downarrow\uparrow\uparrow-\uparrow\downarrow\uparrow)$
				\\

				&	$\ket{\frac{1}{2},-\frac{1}{2}}$ 
				&   ${\frac{1}{\sqrt{2}}}(udc + duc) \otimes {\frac{1}{\sqrt{6}}}(\uparrow\downarrow\downarrow+\downarrow\uparrow\downarrow-2\downarrow\downarrow\uparrow)$
				\\
				\hline
		
				\multirow{2}{*}{$\Sigma_{c}^{0}$}
				&	$\ket{\frac{1}{2},\frac{1}{2}}$  
				&	$ddc \otimes {\frac{1}{\sqrt{6}}}(2\uparrow\uparrow\downarrow-\downarrow\uparrow\uparrow-\uparrow\downarrow\uparrow)$
				\\

				&	$\ket{\frac{1}{2},-\frac{1}{2}}$ 
				&   $ddc \otimes {\frac{1}{\sqrt{6}}}(\uparrow\downarrow\downarrow+\downarrow\uparrow\downarrow-2\downarrow\downarrow\uparrow)$
				\\
				\hline
		
				\multirow{2}{*}{$\Xi_{c}^{\prime +}$ }
				&	$\ket{\frac{1}{2},\frac{1}{2}}$  
				&   ${\frac{1}{\sqrt{2}}}(usc + suc) \otimes {\frac{1}{\sqrt{6}}}(2\uparrow\uparrow\downarrow-\downarrow\uparrow\uparrow-\uparrow\downarrow\uparrow)$
		     	\\

				&	$\ket{\frac{1}{2},-\frac{1}{2}}$ 
				&   ${\frac{1}{\sqrt{2}}}(usc + suc) \otimes {\frac{1}{\sqrt{6}}}(\uparrow\downarrow\downarrow+\downarrow\uparrow\downarrow-2\downarrow\downarrow\uparrow)$
				\\
				\hline
		
				\multirow{2}{*}{$\Xi_{c}^{\prime 0}$ }
				&	$\ket{\frac{1}{2},\frac{1}{2}}$  
				&   ${\frac{1}{\sqrt{2}}}(dsc + sdc) \otimes {\frac{1}{\sqrt{6}}}(2\uparrow\uparrow\downarrow-\downarrow\uparrow\uparrow-\uparrow\downarrow\uparrow)$
				\\
		
				&	$\ket{\frac{1}{2},-\frac{1}{2}}$ 
				&   ${\frac{1}{\sqrt{2}}}(dsc + sdc) \otimes {\frac{1}{\sqrt{6}}}(\uparrow\downarrow\downarrow+\downarrow\uparrow\downarrow-2\downarrow\downarrow\uparrow)$
				\\
				\hline
				
				\multirow{2}{*}{$\Omega_{c}^{0}$}
				&	$\ket{\frac{1}{2},\frac{1}{2}}$  
				&	$ssc \otimes {\frac{1}{\sqrt{6}}}(2\uparrow\uparrow\downarrow-\downarrow\uparrow\uparrow-\uparrow\downarrow\uparrow)$
				\\
		
				&	$\ket{\frac{1}{2},-\frac{1}{2}}$ 
				&   $ssc \otimes {\frac{1}{\sqrt{6}}}(\uparrow\downarrow\downarrow+\downarrow\uparrow\downarrow-2\downarrow\downarrow\uparrow)$
				\\
				\hline
		
				\multirow{2}{*}{$\Xi_{c}^{+}$ }
				&	$\ket{\frac{1}{2},\frac{1}{2}}$  
				&   ${\frac{1}{\sqrt{2}}}(usc - suc) \otimes {\frac{1}{\sqrt{2}}}(\uparrow\downarrow\uparrow-\downarrow\uparrow\uparrow)$
				\\
		
				&	$\ket{\frac{1}{2},-\frac{1}{2}}$ 
				&   ${\frac{1}{\sqrt{2}}}(usc - suc) \otimes {\frac{1}{\sqrt{2}}}(\uparrow\downarrow\downarrow-\downarrow\uparrow\downarrow)$
				\\
				\hline
		
				\multirow{2}{*}{$\Xi_{c}^{0}$ }
				&	$\ket{\frac{1}{2},\frac{1}{2}}$  
				&   ${\frac{1}{\sqrt{2}}}(dsc - sdc) \otimes {\frac{1}{\sqrt{2}}}(\uparrow\downarrow\uparrow-\downarrow\uparrow\uparrow)$
				\\
		
				&	$\ket{\frac{1}{2},-\frac{1}{2}}$ 
				&   ${\frac{1}{\sqrt{2}}}(dsc - sdc) \otimes {\frac{1}{\sqrt{2}}}(\uparrow\downarrow\downarrow-\downarrow\uparrow\downarrow)$
				\\
				\hline
		
				\multirow{2}{*}{$\Lambda_{c}^{+}$ }
				&	$\ket{\frac{1}{2},\frac{1}{2}}$  
				&   ${\frac{1}{\sqrt{2}}}(udc - duc) \otimes {\frac{1}{\sqrt{2}}}(\uparrow\downarrow\uparrow-\downarrow\uparrow\uparrow)$
				\\
		
				&	$\ket{\frac{1}{2},-\frac{1}{2}}$ 
				&   ${\frac{1}{\sqrt{2}}}(udc - duc) \otimes {\frac{1}{\sqrt{2}}}(\uparrow\downarrow\downarrow-\downarrow\uparrow\downarrow)$
				\\
			\toprule[1.0pt]
			\toprule[1.0pt]
			\end{tabular}
		\end{table}		
		
	\renewcommand{\arraystretch}{1.6}
		\begin{table}
			\centering
			\caption{The flavor wave functions $\psi_{flavor}$ and the spin wave functions $\chi_{spin}$ of the $S$-wave anti-charmed mesons. Here, $S$ and $S_3$ are the spin and its third component, while the arrow denotes the third component of the quark spin.}
			\label{tab_4} 
			\begin{tabular}{c|l|l}
			\toprule[1.0pt]
			\toprule[1.0pt]
				Mesons 
				&	$\ket{S,S_3}$  
				&	$\psi_{flavor} \otimes \chi_{spin}$
				\\
				\hline

				$\bar{D}^{0}$
				&	$\ket{0,0}$  
				&	$\bar{c} u \otimes {\frac{1}{\sqrt{2}}}(\uparrow\downarrow-\downarrow\uparrow)$
				\\
				\hline
				$D^{-}$
				&	$\ket{0,0}$  
				&	$\bar{c} d \otimes {\frac{1}{\sqrt{2}}}(\uparrow\downarrow-\downarrow\uparrow)$

				\\
				\hline
				$D_{s}^{-}$
				&	$\ket{0,0}$  
				&	$\bar{c} s \otimes {\frac{1}{\sqrt{2}}}(\uparrow\downarrow-\downarrow\uparrow)$
				\\
				\hline
	
				\multirow{3}{*}{$\bar{D}^{*0}$}
				&	$\ket{1,1}$  
				&	$\bar{c} u \otimes \uparrow\uparrow$
				\\

				&	$\ket{1,0}$  
				&	$\bar{c} u \otimes {\frac{1}{\sqrt{2}}}(\uparrow\downarrow + \downarrow\uparrow)$
				\\

				&	$\ket{1,-1}$  
				&	$\bar{c} u \otimes \downarrow\downarrow$
				\\
				\hline
	
				\multirow{3}{*}{$D^{*-}$}
				&	$\ket{1,1}$  
				&	$\bar{c} d \otimes \uparrow\uparrow$
				\\

				&	$\ket{1,0}$  
				&	$\bar{c} d \otimes {\frac{1}{\sqrt{2}}}(\uparrow\downarrow + \downarrow\uparrow)$
				\\

				&	$\ket{1,-1}$  
				&	$\bar{c} d \otimes \downarrow\downarrow$
				\\
				\hline
	
				\multirow{3}{*}{$D_{s}^{*-}$}
				&	$\ket{1,1}$  
				&	$\bar{c} s \otimes \uparrow\uparrow$
				\\

				&	$\ket{1,0}$  
				&	$\bar{c} s \otimes {\frac{1}{\sqrt{2}}}(\uparrow\downarrow + \downarrow\uparrow)$
				\\

				&	$\ket{1,-1}$  
				&	$\bar{c} s \otimes \downarrow\downarrow$
				\\
			\toprule[1.0pt]
			\toprule[1.0pt]
			\end{tabular}
		\end{table}

		\renewcommand{\arraystretch}{1.65}
		\begin{table}[htbp]
			\centering
			\caption{The expressions for different flavor wave function of the octet hidden-charm molecular pentaquark states under $SU(3)$ symmetry in the molecular model.}
			\label{tab_2} 		
			\begin{tabular}{c|c|c}
				\toprule[1.0pt]
				\toprule[1.0pt]
				States &  Flavor & Wave functions 
				\\
				\hline
				\multirow{2}{*}	{${P_{\psi}^{N^{+}}}$} 
				&	$8_{1f}$
				&	$-\sqrt{\frac{1}{3}}\Sigma_c^{+}\bar{D}^{(*)0}+\sqrt{\frac{2}{3}}\Sigma_{c}^{++}D^{(*)-}$
				\\

				&	$8_{2f}$
				&	$\Lambda_{c}^{+}\Bar{D}^{(*)0}$
				\\
				\hline

				\multirow{2}{*}	{${P_{\psi}^{N^{0}}}$} 
				&	$8_{1f}$
				&	$\sqrt{\frac{1}{3}}\Sigma_c^{+}D^{(*)-}-\sqrt{\frac{2}{3}}\Sigma_{c}^{0}\bar{D}^{(*)0}$
				\\

				&	$8_{2f}$
				&	$\Lambda_{c}^{+}D^{(*)-}$
				\\
				\hline

				\multirow{2}{*}	{${P_{\psi s}^{\Sigma^{+}}}$} 
				&	$8_{1f}$
				&	$\sqrt{\frac{1}{3}}\Xi_c^{\prime+}\bar{D}^{(*)0}-\sqrt{\frac{2}{3}}\Sigma_{c}^{++}D_s^{(*)-}$
				\\

				&	$8_{2f}$
				&	$\Xi_{c}^{+}\Bar{D}^{(*)0}$
				\\
				\hline

				\multirow{2}{*}	{${P_{\psi s}^{\Sigma^{0}}}$} 
				&	$8_{1f}$
				&	$\sqrt{\frac{1}{6}}\Xi_c^{\prime+}D^{(*)-}+\sqrt{\frac{1}{6}}\Xi_{c}^{\prime0}\bar{D}^{(*)0}-\sqrt{\frac{2}{3}}\Sigma_{c}^{+}D_s^{(*)-}$
				\\

				&	$8_{2f}$
				&	$\sqrt{\frac{1}{2}}\Xi_c^{+}D^{(*)-} + \sqrt{\frac{1}{2}}\Xi_{c}^{0}\bar{D}^{(*)0} $
				\\
				\hline

				\multirow{2}{*}	{${P_{\psi s}^{\Lambda^{0}}}$}
				&	$8_{1f}$
				&	$\sqrt{\frac{1}{2}}\Xi_c^{\prime+}D^{(*)-} - \sqrt{\frac{1}{2}}\Xi_{c}^{\prime0}\bar{D}^{(*)0} $
				
				\\

				&	$8_{2f}$
				&	$\sqrt{\frac{1}{6}}\Xi_c^{+}D^{(*)-}-\sqrt{\frac{1}{6}}\Xi_{c}^{0}\bar{D}^{(*)0}-\sqrt{\frac{2}{3}}\Lambda_{c}^{+}D_s^{(*)-}$
				\\
				\hline

				\multirow{2}{*}	{${P_{\psi s}^{\Sigma^{-}}}$} 
				&	$8_{1f}$
				&	$\sqrt{\frac{1}{3}}\Xi_c^{\prime0}D^{(*)-}-\sqrt{\frac{2}{3}}\Sigma_{c}^{0}D_s^{(*)-}$
				\\

				&	$8_{2f}$
				&	$\Xi_{c}^{0}D^{(*)-}$
				\\
				\hline

				\multirow{2}{*}	{${P_{\psi ss}^{N^{0}}}$} 
				&	$8_{1f}$
				&	$\sqrt{\frac{1}{3}}\Xi_c^{\prime+}D_{s}^{(*)-}-\sqrt{\frac{2}{3}}\Omega_{c}^{0}\bar{D}^{(*)0}$
				\\
				%			\cline{2-3}
				
				&	$8_{2f}$
				&	$\Xi_{c}^{+}D_{s}^{(*)-}$
				\\
				\hline

				\multirow{2}{*}	{${P_{\psi ss}^{N^{-}}}$} 
				&	$8_{1f}$
				&	$\sqrt{\frac{1}{3}}\Xi_c^{\prime0}D_{s}^{(*)-}-\sqrt{\frac{2}{3}}\Omega_{c}^{0}D^{(*)-}$
				\\

				&	$8_{2f}$
				&	$\Xi_{c}^{0}D_{s}^{(*)-}$
				\\
			\toprule[1.0pt]
			\toprule[1.0pt]
			\end{tabular}	
		\end{table}

	\section{Magnetic Moments of The octet Hidden-Charm Molecular Pentaquark Family }
	\label{sec3} 
		In this section, we adopt the constituent quark model to obtain the magnetic moment of hadrons. The constituent quark model is extensively applied to study various properties of the hadronic states in the past decades~\cite{Liu:2008qb,Vijande:2004he,Valcarce:2005em}, while the magnetic moments of the hadronic molecular states are focused~\cite{He:2019ify,Karliner:2015ina,Burns:2015dwa,Monemzadeh:2016eoj,Yang:2015bmv,Chen:2016heh,Nakamura:2021dix,Ling:2021lmq}.Magnetic moments of the hadron states are composed of two parts: the spin magnetic moment $\mu_{spin}$ and the orbital magnetic moment $\mu_{orbital}$  
		\begin{eqnarray}
			\mu = \mu_{spin}+ \mu_{orbital}.
		\end{eqnarray}
		In this work, we focus on the $ S $-wave pentaquark states composed of a singly charmed baryon and an anti-charmed meson in the molecular model. Thence, the orbital magnetic moment $\mu_{orbital}$ between baryons and mesons will not be considered in the following calculations. 

		At the quark level, the operators of the magnetic moments are
		\begin{eqnarray}
			\hat{\mu}_{spin} = {\sum_{i}}{\frac{Q_{i}}{2M_{i}}}\hat{\sigma}_{i},
		\end{eqnarray}
		where $Q_i$, $M_i $, and $\hat{\sigma}_{i}$ denote charge, mass, and Pauli’s spin matrix of the $i$th quark, respectively.

		The magnetic moments of the hidden-charm molecular pentaquark states are closely related to the magnetic moments of the singly charmed baryons and the magnetic moments of the anti-charmed mesons. The magnetic moment of the $ S $-wave molecular  pentaquark states include the sum of the baryon spin magnetic moment and the meson spin magnetic moment
		\begin{eqnarray}
			\hat{\mu} = \hat{\mu}_{B} + \hat{\mu}_{M},
		\end{eqnarray}
		where the subscripts $B$ and $M$ represent baryon and meson respectively.  
		At quark level, the magnetic moment of quark can be obtained by the following matrix elements
		\begin{align}
			\bra{q\uparrow}\hat{\mu}_{z}\ket{q\uparrow} 	&= \frac{Q_q}{2M_{q}}, \\
			\bra{q\downarrow}\hat{\mu}_{z}\ket{q\downarrow} 	&= -\frac{Q_q}{2M_{q}},
		\end{align}
		the arrow stands for the third component of the quark spin. As an example, we derive the magnetic moment of the $\Sigma_{c}^{++}$ baryon as follows
		\begin{align}
			\mu_{\Sigma_{c}^{++}} 
			&=
			\left \langle uuc \otimes {\frac{1}{\sqrt{6}}}(2\uparrow\uparrow\downarrow-\downarrow\uparrow\uparrow-\uparrow\downarrow\uparrow)
			\nonumber
			\right.
			\\
			&\left.
			\ \ \ \ \ 
			\left\lvert 
			\hat{\mu}_{spin} 
			\right\lvert	
			uuc \otimes {\frac{1}{\sqrt{6}}}(2\uparrow\uparrow\downarrow-\downarrow\uparrow\uparrow-\uparrow\downarrow\uparrow)
			\right\rangle 
			\nonumber
			\\
			&=
			\frac{4}{3}\mu_{u} - \frac{1}{3}\mu_{c}.
		\end{align}
		
		In Table \ref{tab_5}, we collect the magnetic moment expressions and numerical results for singly charmed baryons and anti-charmed mesons, which are expressed as the combination of the magnetic moments of their constituent quarks. 
		In the numerical analysis, we take the constituent quark masses as input \cite{Wang:2018gpl}:
		\begin{align}
			m_u \ &=\ m_d \ =\  0.336\ \mbox{GeV}, \nonumber
			\\ 	
			m_s \ &=\ 0.540\ \mbox{GeV},\nonumber
			\\  
			m_c \ &=\ 1.660\ \mbox{GeV}. \nonumber	
		\end{align}

		\renewcommand\tabcolsep{0.35cm}
		\renewcommand{\arraystretch}{1.50}
		\begin{table}[t]
			\centering
			\vspace{0.5em}
			\caption{The magnetic moments of the singly charmed baryons and anti-charmed mesons, in unit of the nuclear magnetic moment $\mu_{N}$.}
			\label{tab_5} 
			\begin{tabular}{c|c|c|c}
			\toprule[1.0pt]
			\toprule[1.0pt]
				Flavor & Hadrons & Expressions &  Results
				\\
				\hline

				\multirow{6}{*}{$6_f$}	
				&	$\Sigma_{c}^{++}$ 
				&	${\frac{4}{3}}\mu_{u} - {\frac{1}{3}}\mu_{c}$ 
				&   $2.36$
				\\

				&	$\Sigma_{c}^{+}$ 
				&	${\frac{2}{3}}\mu_{u} + {\frac{2}{3}}\mu_{d} - {\frac{1}{3}}\mu_{c}$ 
				&   $0.49$
				\\

				&	$\Sigma_{c}^{0}$ 
				&	${\frac{4}{3}}\mu_{d} - {\frac{1}{3}}\mu_{c}$ 
				&   $-1.37$
				\\

				&	$\Xi_{c}^{\prime +}$ 
				&	${\frac{2}{3}}\mu_{u} + {\frac{2}{3}}\mu_{s} - {\frac{1}{3}}\mu_{c}$ 
				&   $0.73$
				\\

				&	$\Xi_{c}^{\prime 0}$ 
				&	${\frac{2}{3}}\mu_{d} + {\frac{2}{3}}\mu_{s} - {\frac{1}{3}}\mu_{c}$ 
				&   $-1.13$
				\\

				&	$\Omega_{c}^{0}$ 
				&	${\frac{4}{3}}\mu_{s}-{\frac{1}{3}}\mu_{c}$ 
				&   $-0.90$
				\\
				\hline

				\multirow{3}{*}{$\bar{3}_f$}	
				&	$\Xi_{c}^{+}$ 
				&	$\mu_{c}$ 
				&   $0.38$
				\\
	
				&	$\Xi_{c}^{0}$ 
				&	$\mu_{c}$ 
				&   $0.38$
				\\

				&	$\Lambda_{c}^{+}$ 
				&	$\mu_{c}$ 
				&   $0.38$
				\\
				\hline

				\multirow{3}{*}{$3_f$}	
				&	$\bar{D}^{*0}$ 
				&	$\mu_{u} + \mu_{\bar{c}}$ 
				&   $1.48$
				\\

				&	$D^{*-}$ 
				&	$\mu_{d} + \mu_{\bar{c}}$  
				&   $-1.31$
				\\

				&	$D_{s}^{*-}$ 
				&	$\mu_{s} + \mu_{\bar{c}}$ 
				&   $-0.96$
				\\
			\toprule[1.0pt]
			\toprule[1.0pt]
			\end{tabular}
		\end{table}		
		
		Based on the magnetic moments of the singly charmed	baryons and anti-charmed mesons, we obtain the magnetic moments of the octet hidden-charm molecular pentaquark states. In Table \ref{tab_6}, we present the expressions and numerical results of the magnetic moments of the S-wave octet hidden-charm molecular  pentaquark states.
		\renewcommand{\arraystretch}{2.3}
\begin{table*}[htbp]
	\centering
	\caption{The magnetic moments of the $ S $-wave octet hidden-charm pentaquark family. The  $J_{b} ^{P_{b}}\otimes J_{m} ^{P_{m}} $ are corresponding to the angular momentum and parity of baryon and meson, respectively. The magnetic moment is in unit of the nuclear magnetic moment $\mu_{N}$.}
	\label{tab_6}
	\vspace{0.5em}
	\resizebox{178mm}{!}{	
		\begin{tabular}{c|c|c|cc|cc}
			\toprule[1.0pt]
			\toprule[1.0pt]
			States 
			&   $J_{b} ^{P_{b}}\otimes J_{m} ^{P_{m}} $
			&	$I(J^{P})$ 
			&	Expressions$(8_{1f})$ 
			&	Results
			&	Expressions$(8_{2f})$ 
			&	Results
			\\
			\hline
			\multirow{3}{*}{${P_{\psi}^{N^{+}}}$}
			&	$\frac{1}{2}^{+} \otimes 0^{-}$
			&	${\frac{1}{2}}{(\frac{1}{2})}^{-}$
			&	${\frac{2}{3}}\mu_{\Sigma_{c}^{++}} + {\frac{1}{3}}\mu_{\Sigma_{c}^{+}}$
			&	1.74
			&	$\mu_{\Lambda_{c}^{+}}$
			&	0.38
			\\
			\cline{2-3}
			%%%%%%%%%%%%%%%%%%%%%%%%%%%%%%%%%
			
			&	\multirow{2}{*}{$\frac{1}{2}^{+} \otimes 1^{-}$}
			&	${\frac{1}{2}}{(\frac{1}{2})}^{-}$
			&	$-{\frac{2}{9}}\mu_{\Sigma_{c}^{++}} - {\frac{1}{9}}\mu_{\Sigma_{c}^{+}} + {\frac{4}{9}}\mu_{D^{*-}} + {\frac{2}{9}}\mu_{\bar{D}^{*0}}$
			&	$-0.83$
			&	$-{\frac{1}{3}}\mu_{\Lambda_{c}^{+}} + {\frac{2}{3}}\mu_{\bar{D}^{*0}}$
			&	0.86
			\\
			%%%%%%%%%%%%%%%%%%%%%%%%%%%%%%%%%
			
			&	
			&	${\frac{1}{2}}{(\frac{3}{2})}^{-}$
			&	${\frac{2}{3}}\mu_{\Sigma_{c}^{++}} + {\frac{1}{3}}\mu_{\Sigma_{c}^{+}} + {\frac{2}{3}}\mu_{D^{*-}} + {\frac{1}{3}}\mu_{\bar{D}^{*0}}$
			&	1.36
			&	$\mu_{\Lambda_{c}^{+}} + \mu_{\bar{D}^{*0}}$
			&	1.86
			\\
			\hline
			%%%%%%%%%%%%%%%%%%%%%%%%%%%%%%%%%
			\multirow{3}{*}{${P_{\psi}^{N^{0}}}$}
			&	$\frac{1}{2}^{+} \otimes 0^{-}$
			&	${\frac{1}{2}}{(\frac{1}{2})}^{-}$
			&	${\frac{2}{3}}\mu_{\Sigma_{c}^{0}} + {\frac{1}{3}}\mu_{\Sigma_{c}^{+}}$
			&	$-0.75$
			&	$\mu_{\Lambda_{c}^{+}}$
			&	0.38
			\\
			\cline{2-3}
			%%%%%%%%%%%%%%%%%%%%%%%%%%%%%%%%%
			
			&	\multirow{2}{*}{$\frac{1}{2}^{+} \otimes 1^{-}$}
			&	${\frac{1}{2}}{(\frac{1}{2})}^{-}$
			&	$-{\frac{2}{9}}\mu_{\Sigma_{c}^{0}} - {\frac{1}{9}}\mu_{\Sigma_{c}^{+}} + {\frac{4}{9}}\mu_{\bar{D}^{*0}} + {\frac{2}{9}}\mu_{D^{*-}}$
			&	0.62
			&	$-{\frac{1}{3}}\mu_{\Lambda_{c}^{+}} + {\frac{2}{3}}\mu_{D^{*-}}$
			&	$-1.00$
			\\
			%%%%%%%%%%%%%%%%%%%%%%%%%%%%%%%%%
			
			&	
			&	${\frac{1}{2}}{(\frac{3}{2})}^{-}$
			&	${\frac{2}{3}}\mu_{\Sigma_{c}^{0}} + {\frac{1}{3}}\mu_{\Sigma_{c}^{+}} + {\frac{2}{3}}\mu_{\bar{D}^{*0}} + {\frac{1}{3}}\mu_{D^{*-}}$
			&	-0.19
			&	$\mu_{\Lambda_{c}^{+}} + \mu_{D^{*-}}$
			&	$-0.93$
			\\
			\hline
			%%%%%%%%%%%%%%%%%%%%%%%%%%%%%%%%%
			\multirow{3}{*}{${P_{\psi s}^{\Sigma^{+}}}$}
			&	$\frac{1}{2}^{+} \otimes 0^{-}$
			&	$1{(\frac{1}{2})}^{-}$
			&	${\frac{2}{3}}\mu_{\Sigma_{c}^{++}} + {\frac{1}{3}}\mu_{\Xi_{c}^{\prime +}}$
			&	1.81
			&	$\mu_{\Xi_{c}^{+}}$
			&	0.38
			\\
			\cline{2-3}
			%%%%%%%%%%%%%%%%%%%%%%%%%%%%%%%%%
			
			&	\multirow{2}{*}{$\frac{1}{2}^{+} \otimes 1^{-}$}
			&	$1{(\frac{1}{2})}^{-}$
			&	$-{\frac{2}{9}}\mu_{\Sigma_{c}^{++}} - {\frac{1}{9}}\mu_{\Xi_{c}^{\prime +}} + {\frac{4}{9}}\mu_{D_{s}^{*-}} + {\frac{2}{9}}\mu_{\bar{D}^{*0}}$
			&	$-0.70$
			&	$-{\frac{1}{3}}\mu_{\Xi_{c}^{+}} + {\frac{2}{3}}\mu_{\bar{D}^{*0}}$
			&	0.86
			\\
			%%%%%%%%%%%%%%%%%%%%%%%%%%%%%%%%%
			
			&	
			&	$1{(\frac{3}{2})}^{-}$
			&	${\frac{2}{3}}\mu_{\Sigma_{c}^{++}} + {\frac{1}{3}}\mu_{\Xi_{c}^{\prime +}} + {\frac{2}{3}}\mu_{D_{s}^{*-}} + {\frac{1}{3}}\mu_{\bar{D}^{*0}}$
			&	1.67
			&	$\mu_{\Xi_{c}^{+}} + \mu_{\bar{D}^{*0}}$
			&	1.86
			\\
			\hline
			%%%%%%%%%%%%%%%%%%%%%%%%%%%%%%%%%
			\multirow{3}{*}{${P_{\psi s}^{\Sigma^{0}}}$}
			&	$\frac{1}{2}^{+} \otimes 0^{-}$
			&	${1}{(\frac{1}{2})}^{-}$
			&	${\frac{1}{6}}\mu_{\Xi_{c}^{\prime +}} + {\frac{1}{6}}\mu_{\Xi_{c}^{\prime 0}} + {\frac{2}{3}}\mu_{\Sigma_{c}^{+}}$
			&	0.26
			&	${\frac{1}{2}}\mu_{\Xi_{c}^{+}} + {\frac{1}{2}}\mu_{\Xi_{c}^{0}}$
			&	0.38
			\\
			\cline{2-3}
			%%%%%%%%%%%%%%%%%%%%%%%%%%%%%%%%%
			
			&	\multirow{2}{*}{$\frac{1}{2}^{+} \otimes 1^{-}$}
			&	$1{(\frac{1}{2})}^{-}$
			&	$-{\frac{2}{9}}\mu_{\Sigma_{c}^{+}} - {\frac{1}{18}}\mu_{\Xi_{c}^{\prime +}} - {\frac{1}{18}}\mu_{\Xi_{c}^{\prime 0}} + {\frac{4}{9}}\mu_{D_{s}^{*-}} + {\frac{1}{9}}\mu_{\bar{D}^{*0}} + {\frac{1}{9}}\mu_{D^{*-}}$
			&	$-0.49$
			&	$-{\frac{1}{6}}\mu_{\Xi_{c}^{+}} - {\frac{1}{6}}\mu_{\Xi_{c}^{0}} + {\frac{1}{3}}\mu_{D^{*-}} + {\frac{1}{3}}\mu_{\bar{D}^{*0}}$
			&	$-0.07$
			\\
			%%%%%%%%%%%%%%%%%%%%%%%%%%%%%%%%%
			
			&	
			&	$1{(\frac{3}{2})}^{-}$
			&	${\frac{2}{3}}\mu_{\Sigma_{c}^{+}} + {\frac{1}{6}}\mu_{\Xi_{c}^{\prime +}} + {\frac{1}{6}}\mu_{\Xi_{c}^{\prime 0}} + {\frac{2}{3}}\mu_{D_{s}^{*-}} + {\frac{1}{6}}\mu_{\bar{D}^{*0}} + {\frac{1}{6}}\mu_{D^{*-}}$
			&	$-0.35$
			&	${\frac{1}{2}}\mu_{\Xi_{c}^{+}} + {\frac{1}{2}}\mu_{\Xi_{c}^{0}} + {\frac{1}{2}}\mu_{D^{*-}} + {\frac{1}{2}}\mu_{\bar{D}^{*0}}$
			&	0.47
			\\
			\hline
			%%%%%%%%%%%%%%%%%%%%%%%%%%%%%%%%% 	
			\multirow{3}{*}{${P_{\psi s}^{\Sigma^{-}}}$}
			&	$\frac{1}{2}^{+} \otimes 0^{-}$
			&	$1{(\frac{1}{2})}^{-}$
			&	${\frac{2}{3}}\mu_{\Sigma_{c}^{0}} + {\frac{1}{3}}\mu_{\Xi_{c}^{\prime 0}}$
			&	$-1.29$
			&	$\mu_{\Xi_{c}^{0}}$
			&	0.38
			\\
			\cline{2-3}
			%%%%%%%%%%%%%%%%%%%%%%%%%%%%%%%%%
			
			&	\multirow{2}{*}{$\frac{1}{2}^{+} \otimes 1^{-}$}
			&	$1{(\frac{1}{2})}^{-}$
			&	$-{\frac{2}{9}}\mu_{\Sigma_{c}^{0}} - {\frac{1}{9}}\mu_{\Xi_{c}^{\prime 0}} + {\frac{4}{9}}\mu_{D_{s}^{*-}} + {\frac{2}{9}}\mu_{D^{*-}}$
			&	$-0.29$
			&	$-{\frac{1}{3}}\mu_{\Xi_{c}^{0}} + {\frac{2}{3}}\mu_{D^{*-}}$
			&	$-1.00$
			\\
			%%%%%%%%%%%%%%%%%%%%%%%%%%%%%%%%%
			
			&	
			&	$1{(\frac{3}{2})}^{-}$
			&	${\frac{2}{3}}\mu_{\Sigma_{c}^{0}} + {\frac{1}{3}}\mu_{\Xi_{c}^{\prime 0}} + {\frac{2}{3}}\mu_{D_{s}^{*-}} + {\frac{1}{3}}\mu_{D^{*-}}$
			&	$-2.36$
			&	$\mu_{\Xi_{c}^{0}} + \mu_{D^{*-}}$
			&	$-0.93$
			\\
			\hline
			%%%%%%%%%%%%%%%%%%%%%%%%%%%%%%%%% 
			\multirow{3}{*}{${P_{\psi s}^{\Lambda^{0}}}$}
			&	$\frac{1}{2}^{+} \otimes 0^{-}$
			&	$0{(\frac{1}{2})}^{-}$
			&	${\frac{1}{2}}\mu_{\Xi_{c}^{\prime +}} + {\frac{1}{2}}\mu_{\Xi_{c}^{\prime 0}}$
			&	$-0.20$
			&	${\frac{1}{6}}\mu_{\Xi_{c}^{+}} + {\frac{1}{6}}\mu_{\Xi_{c}^{0}} + {\frac{2}{3}}\mu_{\Lambda_{c}^{+}}$
			&	0.38
			\\
			\cline{2-3}
			%%%%%%%%%%%%%%%%%%%%%%%%%%%%%%%%%
			
			&	\multirow{2}{*}{$\frac{1}{2}^{+} \otimes 1^{-}$}
			&	$0{(\frac{1}{2})}^{-}$
			&	$-{\frac{1}{6}}\mu_{\Xi_{c}^{\prime +}} - {\frac{1}{6}}\mu_{\Xi_{c}^{\prime 0}} + {\frac{1}{3}}\mu_{D^{*-}} + {\frac{1}{3}}\mu_{\bar{D}^{*0}}$
			&	0.13
			&	$-{\frac{2}{9}}\mu_{\Lambda_{c}^{+}} - {\frac{1}{18}}\mu_{\Xi_{c}^{+}} - {\frac{1}{18}}\mu_{\Xi_{c}^{0}} + {\frac{4}{9}}\mu_{D_{s}^{*-}} + {\frac{1}{9}}\mu_{\bar{D}^{*0}} + {\frac{1}{9}}\mu_{D^{*-}}$
			&	$-0.53$
			\\
			%%%%%%%%%%%%%%%%%%%%%%%%%%%%%%%%%
			
			&	
			&	$0{(\frac{3}{2})}^{-}$
			&	${\frac{1}{2}}\mu_{\Xi_{c}^{\prime +}} + {\frac{1}{2}}\mu_{\Xi_{c}^{\prime 0}} + {\frac{1}{2}}\mu_{D^{*-}} + {\frac{1}{2}}\mu_{\bar{D}^{*0}}$
			&	$-0.11$
			&	${\frac{2}{3}}\mu_{\Lambda_{c}^{+}} + {\frac{1}{6}}\mu_{\Xi_{c}^{+}} + {\frac{1}{6}}\mu_{\Xi_{c}^{0}} + {\frac{2}{3}}\mu_{D_{s}^{*-}} + {\frac{1}{6}}\mu_{\bar{D}^{*0}} + {\frac{1}{6}}\mu_{D^{*-}}$
			&	$-0.23$
			\\
			\hline
			
			%%%%%%%%%%%%%%%%%%%%%%%%%%%%%%%%% 	 
			\multirow{3}{*}{ ${P_{\psi ss}^{N^{0}}}$}
			&	$\frac{1}{2}^{+} \otimes 0^{-}$
			&	${\frac{1}{2}}{(\frac{1}{2})}^{-}$
			&	${\frac{2}{3}}\mu_{\Omega_{c}^{0}} + {\frac{1}{3}}\mu_{\Xi_{c}^{\prime +}}$
			&	$-0.36$
			&	$\mu_{\Xi_{c}^{+}}$
			&	0.38
			\\
			\cline{2-3}
			%%%%%%%%%%%%%%%%%%%%%%%%%%%%%%%%%
			
			&	\multirow{2}{*}{$\frac{1}{2}^{+} \otimes 1^{-}$}
			&	${\frac{1}{2}}{(\frac{1}{2})}^{-}$
			&	$-{\frac{2}{9}}\mu_{\Omega_{c}^{0}} - {\frac{1}{9}}\mu_{\Xi_{c}^{\prime +}} + {\frac{4}{9}}\mu_{\bar{D}^{*0}} + {\frac{2}{9}}\mu_{D_{s}^{*-}}$
			&	0.57
			&	$-{\frac{1}{3}}\mu_{\Xi_{c}^{+}} + {\frac{2}{3}}\mu_{D_{s}^{*-}}$
			&	$-0.76$
			\\
			%%%%%%%%%%%%%%%%%%%%%%%%%%%%%%%%%
			
			&	
			&	${\frac{1}{2}}{(\frac{3}{2})}^{-}$
			&	${\frac{2}{3}}\mu_{\Omega_{c}^{0}} + {\frac{1}{3}}\mu_{\Xi_{c}^{\prime +}} + {\frac{2}{3}}\mu_{\bar{D}^{*0}} + {\frac{1}{3}}\mu_{D_{s}^{*-}}$
			&	0.32
			&	$\mu_{\Xi_{c}^{+}} + \mu_{D_{s}^{*-}}$
			&	$-0.58$
			\\
			\hline
			%%%%%%%%%%%%%%%%%%%%%%%%%%%%%%%%% 
			\multirow{3}{*}{${P_{\psi ss}^{N^{-}}}$}
			&	$\frac{1}{2}^{+} \otimes 0^{-}$
			&	${\frac{1}{2}}{(\frac{1}{2})}^{-}$
			&	${\frac{2}{3}}\mu_{\Omega_{c}^{0}} + {\frac{1}{3}}\mu_{\Xi_{c}^{\prime 0}}$
			&	$-0.98$
			&	$\mu_{\Xi_{c}^{0}}$
			&	0.38
			\\
			\cline{2-3}
			%%%%%%%%%%%%%%%%%%%%%%%%%%%%%%%%%
			
			&	\multirow{2}{*}{$\frac{1}{2}^{+} \otimes 1^{-}$}
			&	${\frac{1}{2}}{(\frac{1}{2})}^{-}$
			&	$-{\frac{2}{9}}\mu_{\Omega_{c}^{0}} - {\frac{1}{9}}\mu_{\Xi_{c}^{\prime 0}} + {\frac{4}{9}}\mu_{D^{*-}} + {\frac{2}{9}}\mu_{D_{s}^{*-}}$
			&	$-0.47$
			&	$-{\frac{1}{3}}\mu_{\Xi_{c}^{0}} + {\frac{2}{3}}\mu_{D_{s}^{*-}}$
			&	$-0.76$
			\\
			%%%%%%%%%%%%%%%%%%%%%%%%%%%%%%%%%
			
			&	
			&	${\frac{1}{2}}{(\frac{3}{2})}^{-}$
			&	${\frac{2}{3}}\mu_{\Omega_{c}^{0}} + {\frac{1}{3}}\mu_{\Xi_{c}^{\prime 0}} + {\frac{2}{3}}\mu_{D^{*-}} + {\frac{1}{3}}\mu_{D_{s}^{*-}}$
			&	$-2.17$
			&	$\mu_{\Xi_{c}^{0}} + \mu_{D_{s}^{*-}}$
			&	$-0.58$
			\\
			\toprule[1.0pt]
			\toprule[1.0pt]
	\end{tabular}}
\end{table*}		

		\begin{figure*}
	\centering
	\includegraphics[width=1\linewidth, height=0.4\textheight]{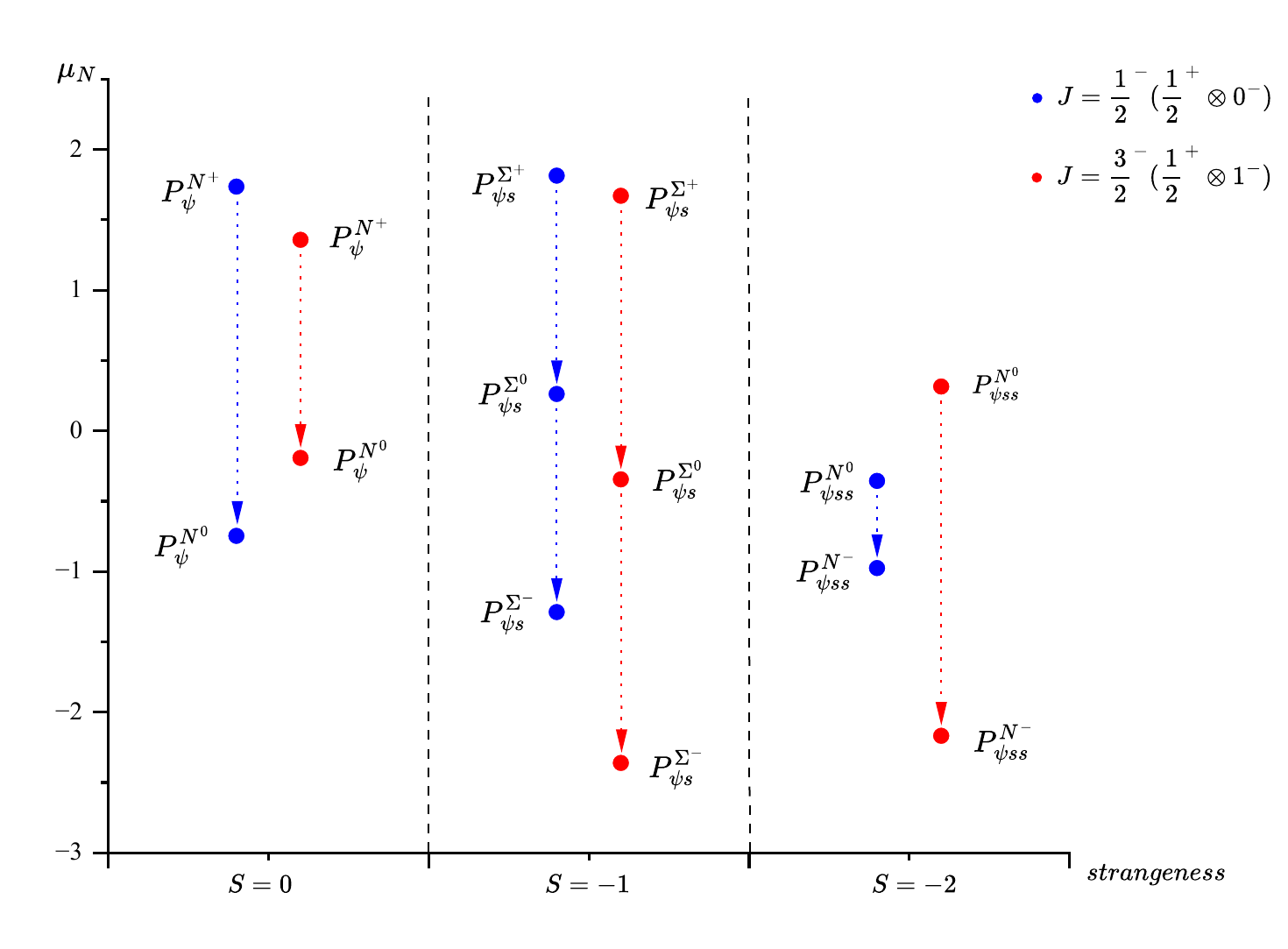}
	\caption{The magnetic moments of the $ S $-wave hidden-charm molecular pentaquark states with the $ 8_{1f} $ flavor representation. Here, the blue arrow points to the lower magnetic moment  in the hidden-charm molecular pentaquark states with quantum number $ J=\frac{1}{2}^{-}(\frac{1}{2}^{+}\otimes0^{-}) $, and the red arrow points to the lower magnetic moment  in the hidden-charm molecular pentaquark states with quantum number $ J=\frac{3}{2}^{-}(\frac{1}{2}^{+}\otimes1^{-}) $.}
	\label{fig:81f}
\end{figure*}
\begin{figure*}
	\centering
	\includegraphics[width=1\linewidth, height=0.4\textheight]{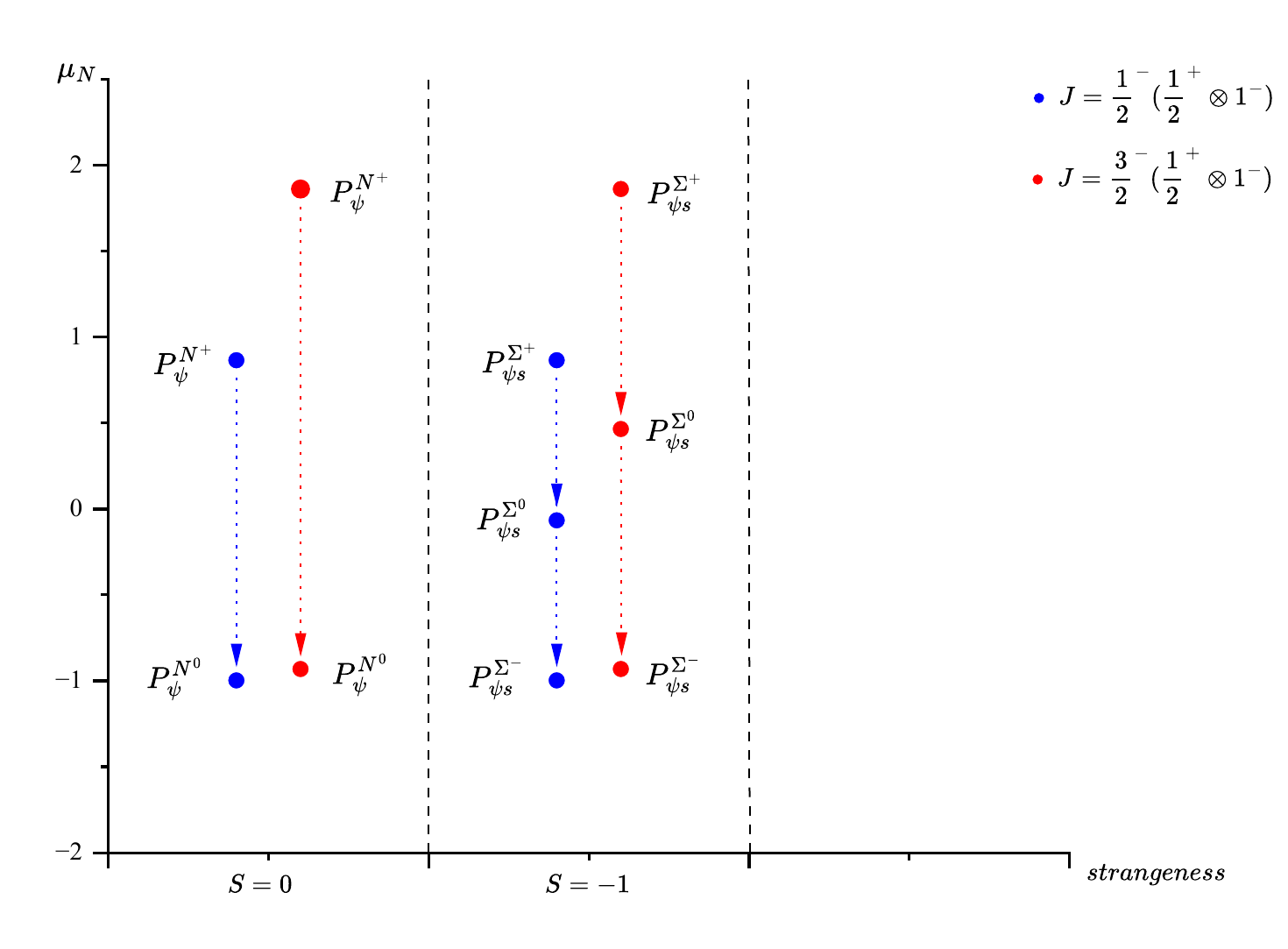}
	\caption{The magnetic moments of the $ S $-wave hidden-charm molecular pentaquark states with the $ 8_{2f} $ flavor representation. Here, the blue arrow points to the lower magnetic moment  in the hidden-charm molecular pentaquark states with quantum number $ J=\frac{1}{2}^{-}(\frac{1}{2}^{+}\otimes1^{-}) $, and the red arrow points to the lower magnetic moment  in the hidden-charm molecular pentaquark states with quantum number $ J=\frac{3}{2}^{-}(\frac{1}{2}^{+}\otimes1^{-}) $.}
	\label{fig:82f}
\end{figure*}
		
	The magnetic moment expressions of the hidden-charm molecular pentaquark family are composed of the magnetic moments of singly charmed baryons and anti-charmed mesons. Different angular momentum couplings have different Clebsch-Gordan coefficients to distinguish them. The magnetic moment corresponding to the same $S$-wave molecular pentaquark states in the  $ 8_{1f} $ flavor representation and $8_{2f} $ flavor representation is completely different. In the $ 8_{2f} $ flavor representation, the magnetic moments of  pentaquark states with spin configuration $J^{P}=\frac{1}{2}^{-}(\frac{1}{2}^{+}\otimes0^{-})$ are all equal to $\mu_{c}=0.38\mu_{N}$. We also notice that the smaller the number of charges, the smaller  magnetic moments of octet  hidden-charm molecular pentaquark states with the same strangeness number. For a more intuitive expression, we have shown the magnetic moments of  the $ S $-wave molecular pentaquark states with  $ 8_{1f} $ flavor representation and $8_{2f} $ flavor representation in Fig. \ref{fig:81f} and Fig. \ref{fig:82f}, respectively. 
	In particular, the Coleman-Glashow sum-rule for the magnetic moments of the octet baryons turns out to hold also for the magnetic moments of the same spin flavor configuration pentaquark states.
\begin{eqnarray}
	\mu_{P1_{\psi}^{N+}}-
	\mu_{P1_{\psi}^{N0}}+\mu_{P1_{\psi s}^{\Sigma-}}-
	\mu_{P1_{\psi s}^{\Sigma+}}
	+\mu_{P1_{\psi ss}^{N0}}-
	\mu_{P1_{\psi ss}^{N-}}=0. \nonumber
\end{eqnarray}
The rules that the magnetic moments satisfy provide us with new directions for exploring the inner structures of hidden-charm pentaquark states.

	\section{The Axial Charge of the octet Hidden-Charm Molecular Pentaquark Family }
	\label{sec4} 

		The axial charge plays a fundamental role in both the electroweak and strong interactions within the standard model. Moreover, it serves as an essential indicator of the spontaneous breaking of chiral symmetry in non-perturbative QCD. The exploration of axial charges has remained continuous endeavor~\cite{Bijnens:1984ec,Jenkins:1991es,Zhu:2000zf,Zhu:2002tn,Choi:2010ty,Deng:2021gnb,Dahiya:2023izc}. Examining the axial charges of the newly discovered hidden-charm pentaquark states presents an effective approach to gain a deeper understanding of these phenomena.
		
At the quark level, the pion-quark interaction reads
		\begin{align}
			\mathscr{L}_{quark}
			&=\frac{1}{2} g_{q} \bar \psi_q\gamma^{\mu}\gamma_5
			\partial_{\mu}\phi \psi_q\sim \frac{1}{2}g_{q} \bar \psi_q \sigma_z\partial_z\phi \psi_q \nonumber\\
			&= \frac{1}{2}\frac{g_q}{f_\pi} (\bar u
			\sigma_z\partial_z\pi_0 u-\bar d \sigma_z\partial_z\pi_0 d)\nonumber\\
			&+\frac{\sqrt{2}}{2}\frac{g_q}{f_{\pi}}(\bar{u}\sigma_z\partial_{z}\pi^{+}d 
			+ \bar{d}\sigma_z\partial_{z}\pi^{-}u),
		\end{align}
		where $g_q$ is the coupling constant at the quark level, $\sigma_z$ is the Pauli matrix. The pion decay constant $f_\pi=92 \mbox{MeV}$. $\phi$ is the pseudoscalar meson field,
		\begin{align} 
			\phi= \sum_{i = 1}^{3}\tau_{i}\phi_{i} 
			\equiv\left(
			\begin{array}{cc}
				\pi_0 &\sqrt{2}\pi^{+}\\
				\sqrt{2}\pi^{-} &-\pi_0\\
			\end{array}
			\right).
		\end{align}

		The nucleon-pion Lagrangian at hadron level reads
		\begin{equation} 
			\mathscr{L}^{N}_{eff}=\frac{1}{2}g_{A} \bar N \gamma^{\mu} \gamma_5 \partial_{\mu}\phi N \sim \frac{g_{A}}{f_\pi}\bar N\frac{\Sigma_{Nz}}{2}\partial_z \pi_0 N,\label{equ:9}
		\end{equation}	
		where $g_{A}$ is the axial charge of the nucleon.
		%$\frac{\Sigma_{Nz}}{2}$ is the $ z $ component of the proton spin operator.
		\begin{align}
			\langle  N, j_3=\frac{1}{2}; ~\pi_0|\mathscr{L}^{N}_{eff} |N, j_3=\frac{1}{2}\rangle = \frac{1}{2}\frac{q_z}{f_\pi}g_{A},\label{equ:10}
		\end{align}
		where $q_z$ is the external momentum of $\pi_0$. At the quark level,
		\begin{equation} 
			\langle  N , j_3=+\frac{1}{2} ;~\pi_0|	\mathscr{L}_{quark}|N, j_3=+\frac{1}{2}\rangle=\frac{5}{6}  \frac{q_z}{f_\pi}g_q.\label{equ:11}
		\end{equation}
		From Eq. (\ref{equ:10}) and Eq. (\ref{equ:11}), we obtain $g_q = \frac{3}{5}g_{A}$  under the quark-hadron duality space.

    The $SU(3)$ invariant Lagrangian of pentaquark state for $J^{P}=\frac{1}{2}^{-}(\frac{1}{2}^{+}\otimes0^{-})$ in $8_{1f}$ flavor representation reads
		\begin{align} 
			\mathscr{L}^{\frac{1}{2}}_{1}
			&=Tr\Big(g_{1}\bar{P} \gamma_{\mu}\gamma^{5}\{\partial^{\mu}\Phi,P\} +f_{1}\bar{P}\gamma_{\mu}\gamma^{5}[\partial^{\mu}\Phi,P]\Big),\label{equ:13}	
		\end{align}
	where $g_{1}$ and $f_{1}$ are independent axial coupling constants of pentaquark state for $J^{P}=\frac{1}{2}^{-}(\frac{1}{2}^{+}\otimes0^{-})$ in $8_{1f}$ flavor representation. $\Phi$ represents the pseudoscalar meson field in $SU(3)$ flavor symmetry
		\begin{align} 
			\Phi = \sum_{a = 1}^{8}\lambda_{a}\Phi_{a} \equiv
			\left(
			\begin{array}{ccc}
				\pi_{0} + \frac{1}{\sqrt{3}}\eta
				&\sqrt{2}\pi^{+}
				&\sqrt{2}K^{+}
				\\
				\sqrt{2}\pi^{-}
				&-\pi_0 + \frac{1}{\sqrt{3}}\eta
				&\sqrt{2}K^{0}
				\\	
				\sqrt{2}K^{-}
				&\sqrt{2}\bar{K}^{0}
				&-\frac{2}{\sqrt{3}}\eta
			\end{array}
			\right).
		\end{align}
		$P$ represents the octet hidden-charm molecular pentaquark states.
		\begin{equation} 
			\setlength{\arraycolsep}{0.1pt}
			P= 
			\left(		
			\begin{array}{ccc}
				\frac{1}{\sqrt{2}}{{}P_{\psi s}^{\Sigma^{0}}}+\frac{1}{\sqrt{6}}{{}P_{\psi s}^{\Lambda^{0}}}
				&{{}P_{\psi s}^{\Sigma^{+}}}
				&{{}P_{\psi}^{N^{+}}}
				\\
				{{}P_{\psi s}^{\Sigma^{-}}}
				&-\frac{1}{\sqrt{2}}{{}P_{\psi s}^{\Sigma^{0}}}+\frac{1}{\sqrt{6}}{{}P_{\psi s}^{\Lambda^{0}}}
				&{{}P_{\psi}^{N^{0}}}
				\\
				{{}P_{\psi ss}^{N^{-}}}
				&{{}P_{\psi ss}^{N^{+}}}
				&\frac{2}{\sqrt{3}}{{}P_{\psi s}^{\Lambda^{0}}}
				\\
			\end{array}
			\right).\label{equ:20}
		\end{equation}
		After bringing the matrix representation of $\Phi$ and $P$ into Eq. (\ref{equ:13}) and keep $\pi_{0}$ meson term, we obtain
	\begin{align}
		\mathscr{L}^{\frac{1}{2}}_{1}
		&=
		{\frac{2}{\sqrt{3}}}{\frac{1}{f_{\pi}}} g_{1} ({{}\bar{P}_{\psi s}^{\Sigma^{0}}}\Sigma_{z}\partial_{z}{\pi^0}{{}P_{\psi s}^{\Lambda^{0}}}
		+
		{{}\bar{P}_{\psi s}^{\Lambda^{0}}}\Sigma_{z}\partial_{z}{\pi^0}{{}P_{\psi s}^{\Sigma^{0}}} )
		\nonumber
		\\
		&-
		{\frac{1}{f_{\pi}}}(f_{1}+g_{1}){{}\bar{P}_{\psi}^{N}}^{0}\Sigma_{z}\partial_{z}{\pi^0}{{}P_{\psi}^{N}}^{0}
		\nonumber\\
		&+
		{\frac{1}{f_{\pi}}}(g_{1}+f_{1}){{}\bar{P}_{\psi}^{N}}^{+}\Sigma_{z}\partial_{z}{\pi^0}{{}P_{\psi}^{N}}^{+}
		\nonumber\\
		&+
		{\frac{1}{f_{\pi}}}(g_{1}-f_{1}){{}\bar{P}_{\psi ss}^{N^{-}}}\Sigma_{z}\partial_{z}{\pi^0}{{}P_{\psi ss}^{N^{-}}}
		\nonumber\\
		&+
		{\frac{1}{f_{\pi}}}(f_{1}-g_{1}){{}\bar{P}_{\psi ss}^{N^{0}}}\Sigma_z\partial_{z}{\pi^0}{{}P_{\psi ss}^{N^{0}}}
		\nonumber\\
		&+
		{\frac{1}{f_{\pi}}}2f_{1}{{}\bar{P}_{\psi s}^{\Sigma^{+}}}\Sigma_z\partial_{z}{\pi^0}{{}P_{\psi s}^{\Sigma^{+}}}
		\nonumber\\
		&-
		{\frac{1}{f_{\pi}}}2f_{1}{{}\bar{P}_{\psi s}^{\Sigma}}^{-}\Sigma_z\partial_{z}{\pi^0}{{}P_{\psi  s}^{\Sigma^{-}}}.\label{equ:21}
	\end{align}
		 We can obtain the axial coupling constant of the octet hidden-charm molecular pentaquark state with the coupling constants $g_{1}$ and $f_{1}$. While Eq. (\ref{equ:21}) solely presents the components associated with the decay of the $\pi_{0}$ meson, the axial coupling constants for decays involving other mesons can also be derived from the constants $g_{1}$ and $f_{1}$.

		 To determine the coupling constants $g_{1}$ and $f_{1}$, similar to the procedure employed for the nucleon, we consider the $\pi_{0}$ meson decay of $P_{\psi}^{N^{+}}$ and ${P_{\psi ss}^{N^{0}}}$ in Eq. (\ref{equ:21}). The Lagrangian for the $\pi_{0}$ decay of ${{}P_{\psi}^{N^{+}}}$ with the spin configuration $J^{P}=\frac{1}{2}^{-}(\frac{1}{2}^{+}\otimes0^{-})$ at the hadron level reads
		\begin{align} 
			\mathscr{L}^{\frac{1}{2}}_{P_{\psi}^{N^{+}}}
			&=
			\frac{1}{2}(g_{1}+f_{1}) {{}\bar{P}_{\psi}^{N^{+}}} \gamma^{\mu} \gamma_5  \partial_{\mu}\Phi{{}P_{\psi}^{N^{+}}}
			\nonumber  \\
			&\sim %约等号
			\frac{g_{1}+f_{1}}{f_\pi}{{}\bar{P}_{\psi}^{N^{+}}}\frac{\Sigma_z}{2}\partial_{z}  {\pi_{0}}{{}P_{\psi}^{N^{+}}}. 
		\end{align}
	The Lagrangian for ${P_{\psi ss}^{N^{0}}}$ with the spin configuration $J^{P}=\frac{1}{2}^{-}(\frac{1}{2}^{+}\otimes0^{-})$ reads
		\begin{align} 
			\mathscr{L}^{\frac{1}{2}}_{{P_{\psi ss}^{N^{0}}}}
			&=
			\frac{1}{2}(f_{1}-g_{1}) {{}\bar{P}_{\psi ss}^{N^{0}}} \gamma^{\mu} \gamma_5 \partial_{\mu}\Phi{{}P_{\psi  ss}^{N^{0}}} 
			\nonumber  \\
			&\sim %约等号
			\frac{f_{1}-g_{1}}{f_\pi}{{}\bar{P}_{\psi ss}^{N^{0}}}\frac{\Sigma_z}{2}\partial_{z} {\pi_{0}}{{}P_{\psi  ss}^{N}},
		\end{align}
		$\frac{\Sigma_z}{2}$ is the spin operator of the hidden-charm pentaquark states. At the hadron level, the axial charges read
		\begin{align} 	
			&\left \langle 
			{{}P_{\psi}^{N}}^{+}; \pi_{0}
			|
			\frac{g_{1}+f_{1}}{f_\pi}{{}\bar{P}_{\psi}^{N}}^{+}\frac{\Sigma_z}{2}\partial_{z}  {\pi_{0}}{{}P_{\psi}^{N}}^{+} 
			|
			{P_{\psi}^{N}}^{+} 
			\right\rangle 
			\nonumber\\
			&= 
			\frac{g_{1}+f_{1}}{2}\frac{q_z}{f_\pi}.	
			\\	
			&\langle 
			{{}P_{\psi ss}^{N^{0}}}; \pi_{0}
			|
			\frac{f_{1}-g_{1}}{f_\pi}{{}\bar{P}_{\psi ss}^{N^{0}}}\frac{\Sigma_z}{2}\partial_{z} {\pi_{0}}{{}P_{\psi  ss}^{N^{0}}} 
			|
			{{}P_{\psi ss}^{N^{0} }}
			\rangle 
			\nonumber\\
			&= 
			\frac{f_{1}-g_{1}}{2}\frac{q_z}{f_\pi}.
		\end{align}
		At the quark level, the axial charge of ${P_{\psi}^{N^{+}}}$ and ${P_{\psi ss}^{N^{0}}}$ with the spin configuration $J^{P}=\frac{1}{2}^{-}(\frac{1}{2}^{+}\otimes0^{-})$ read
		\begin{align} 
			\langle {{}P_{\psi}^{N^{+}}} , +\frac{1}{2} ;~\pi_0| \mathscr{L}_{quark}|{{}P_{\psi}^{N^{+}}},
			+\frac{1}{2}\rangle 
			=\frac{4}{9}\frac{q_z}{f_\pi}g_q,
			\\
			\langle {{}P_{\psi ss}^{N^{0}}} , +\frac{1}{2} ;~\pi_0| \mathscr{L}_{quark}|{{}P_{\psi ss}^{N^{0}}},
			+\frac{1}{2}\rangle
			=\frac{1}{9} \frac{q_z}{f_\pi}g_q.\label{equ:28}
		\end{align}
	Compare with the axial charge of the nucleon, 
		\begin{align} 
			\frac{\frac{1}{2}g_{A}}{\frac{5}{6}g_{q}}	 
			&= 
			\frac{\frac{g_{1}+f_{1}}{2}}{\frac{4}{9}g_{q}} 
			=
			\frac{\frac{f_{1}-g_{1}}{2}}{\frac{1}{9}g_{q}}. \label{equ:28}
		\end{align}
	We obtain $f_{1} = \frac{1}{3}g_{A}$ and $g_{1} = \frac{1}{5}g_{A}$. Similarly, we can obtain the axial charge of the pentaquark states with other flavor-spin configurations. 
	
	The Lagrangian of pentaquark state for $J^{P}=\frac{1}{2}^{-}(\frac{1}{2}^{+}\otimes1^{-})$ in $8_{1f}$ flavor representation reads
		\begin{align} 
			\mathscr{L}^{\frac{1}{2}}_{2}
			&=Tr\Big(g_{2}\bar{P} \gamma_{\mu}\gamma^{5}\{\partial^{\mu}\Phi,P\}
			+f_{2}\bar{P}\gamma_{\mu}\gamma^{5}[\partial^{\mu}\Phi,P]\Big).
		\end{align}
		The Lagrangian of pentaquark state for $J^{P}=\frac{3}{2}^{-}(\frac{1}{2}^{+}\otimes1^{-})$ in $8_{1f}$ flavor representation reads
		\begin{align} 
			\mathscr{L}^{\frac{3}{2}}_{3}
			&=Tr\Big(g_{3}\bar{P}^{\nu} \gamma_{\mu}\gamma^{5}\{\partial^{\mu}\Phi,P_{\nu}\}
			+f_{3}\bar{P}^{\nu}\gamma_{\mu}\gamma^{5}[\partial^{\mu}\Phi,P_{\nu}]\Big).
		\end{align}
		The Lagrangian of pentaquark state for $J^{P}=\frac{1}{2}^{-}(\frac{1}{2}^{+}\otimes0^{-})$ in $8_{2f}$ flavor representation reads
		\begin{align} 
			\mathscr{L}^{\frac{1}{2}}_{4}
			&=Tr\Big(g_{4}\bar{P} \gamma_{\mu}\gamma^{5}\{\partial^{\mu}\Phi,P\}
			+f_{4}\bar{P}\gamma_{\mu}\gamma^{5}[\partial^{\mu}\Phi,P]\Big).
		\end{align}
		The Lagrangian of pentaquark state for $J^{P}=\frac{1}{2}^{-}(\frac{1}{2}^{+}\otimes1^{-})$ in $8_{2f}$ flavor representation reads
		\begin{align} 
			\mathscr{L}^{\frac{1}{2}}_{5}
			&=Tr\Big(g_{5}\bar{P}\gamma_{\mu}\gamma^{5}\{\partial^{\mu}\Phi,P\}
			+f_{5}\bar{P}\gamma_{\mu}\gamma^{5}[\partial^{\mu}\Phi,P]\Big).
		\end{align}
		The Lagrangian of pentaquark state for $J^{P}=\frac{3}{2}^{-}(\frac{1}{2}^{+}\otimes1^{-})$ in $8_{2f}$ flavor representation reads
		\begin{align} 
			\mathscr{L}^{\frac{3}{2}}_{6}
			&=Tr\Big(g_{6}\bar{P}^{\nu}\gamma_{\mu}\gamma^{5}\{\partial^{\mu}\Phi,P_{\nu}\}
			+f_{6}\bar{P}^{\nu}\gamma_{\mu}\gamma^{5}[\partial^{\mu}\Phi,P_{\nu}]\Big).
		\end{align}
	
		\renewcommand{\arraystretch}{1.6}
	\begin{table}[htp]
		\centering
		\caption{Coupling constants $ f _{i}$ and $ g_{i} $ of the pentaquark states with different flavor-spin configurations}
		\label{tab_25} 
		\vspace{0.5em}
		\begin{tabular}{c|c|c|c}
			\toprule[1.0pt]
			\toprule[1.0pt]
			Constants
			&  Values
			& Constants
			&	Values
			\\
			\hline
			$f_{1}$
			& $\frac{1}{3}g_{A}$ 
			& $g_{1}$
			& $\frac{1}{5}g_{A}$
			\\
			\cline{1-4}
			$f_{2}$
			& $-\frac{2}{45}g_{A}$
			& $g_{2}$
			& $-\frac{12}{45}g_{A}$
			\\
			\cline{1-4}
			$f_{3}$
			&  $\frac{13}{90}g_{A}$
			&  $g_{3}$
			&  $-\frac{1}{30}g_{A}$
			\\
			\hline
			$f_{4}$
			&  $0$
			&  $g_{4}$
			&  $0$
			\\
			\cline{1-4}
			$f_{5}$
			&  $\frac{1}{5}g_{A}$ 
			&  $g_{5}$
			&  $\frac{1}{5}g_{A}$ 
			\\
			\cline{1-4}
			$f_{6}$
			&   $\frac{1}{10}g_{A}$ 
			&  $g_{6}$
			&  $\frac{1}{10}g_{A}$ 
			\\
			\toprule[1.0pt]
			\toprule[1.0pt]
		\end{tabular}
	\end{table}

	The numerical results for the coupling constants $ f _{i}$ and $ g_{i} $ of pentaquark states with different flavor-spin configurations are presented in Table \ref{tab_25}.	The axial charge of the octet hidden-charm molecular family in $8_{1f}$ and $8_{2f}$ flavor representations are listed in Table \ref{tab_23} and Table \ref{tab_24} respectively.
	
		\renewcommand{\arraystretch}{1.8}
		\begin{table*}[htbp]
			\centering
			\caption{The axial charges of the octet hidden-charm pentaquark family in $8_{1f}$ flavor representation. The  $J_{b} ^{P_{b}}\otimes J_{m} ^{P_{m}} $ are corresponding to the angular momentum and parity of baryon and meson, respectively.}
			\label{tab_23} 
			\vspace{0.5em}
			\resizebox{180mm}{!}{	
				\begin{tabular}{ c|c|c|c|c|c}
				\toprule[1.0pt]
				\toprule[1.0pt]
					Couplings & Coefficients &Wave functions & $J_{b} ^{P_{b}}\otimes J_{m} ^{P_{m}} $ & $I(J^{P})$ & Results 
					\\
					\hline 
					\multirow{3}{*}{${P_{\psi}^{N^{+}}}{P_{\psi}^{N^{+}}}{\pi_{0}}$}
					&	$f_{1}+g_{1}$ %\multirow{3}{*}{$f+g$}
					& \multirow{3}{*}{${P_{\psi}^{N^{+}}}:-\sqrt{\frac{1}{3}}\Sigma_{c}^{+}\bar{D}^{(*)0}+\sqrt{\frac{2}{3}}\Sigma_{c}^{++}D^{(*)-}$}
					&	$\frac{1}{2}^{+}\times0^{-}$
					&	$\frac{1}{2}(\frac{1}{2})^{-}$
					&	$\frac{8}{15}g_A$
					\\
					\cline{2-2}
					\cline{4-6}

					&	$f_{2}+g_{2}$
					&		
					&	\multirow{2}{*}{$\frac{1}{2}^{+}\times1^{-}$}
					&	$\frac{1}{2}(\frac{1}{2})^{-}$
					&	$-\frac{14}{45}g_A$
					\\
					\cline{2-2}
					\cline{5-6}

					&	$f_{3}+g_{3}$	
					&
					&	
					&	$\frac{1}{2}(\frac{3}{2})^{-}$		
					&	$\frac{1}{9}g_A$
					\\
					\hline					
					
					\multirow{3}{*}{${P_{\psi}^{N^{0}}}{P_{\psi}^{N^{0}}}\pi_{0}$}
					&	$-g_{1}-f_{1}$
					&				\multirow{3}{*}{${P_{\psi}^{N^{0}}}: \sqrt{\frac{1}{3}}\Sigma_{c}^{+}D^{(*)-}-\sqrt{\frac{2}{3}}\Sigma_{c}^{0}\bar{D}^{(*)0}$}
					&	$\frac{1}{2}^{+}\times0^{-}$
					&	$\frac{1}{2}(\frac{1}{2})^{-}$
					&	$-\frac{8}{15}g_A$
					\\
					\cline{2-2}
					\cline{4-6}

					&	$-g_{2}-f_{2}$
					&		
					&	\multirow{2}{*}{$\frac{1}{2}^{+}\times1^{-}$}
					&	$\frac{1}{2}(\frac{1}{2})^{-}$
					&	$\frac{14}{45}g_A$
					\\
					\cline{2-2}
					\cline{5-6}

					&	$-g_{3}-f_{3}$
					&
					&	
					&	$\frac{1}{2}(\frac{3}{2})^{-}$
					&	$-\frac{1}{9}g_A$
					\\
					\hline
					%%%%%%%%%%%%%%%%%%%%%%%%%%%%%%%%%%%%%%%%%%%%%%%%%%%%%%%%%%%%%%%%%%%%%%%%%%%	
					\multirow{3}{*}{${P_{\psi s}^{\Sigma^{+}}}{P_{\psi s}^{\Sigma^{+}}}\pi_{0}$}
					&	$2f_{1}$
					&	\multirow{3}{*}{${P_{\psi s}^{\Sigma^{+}}}:\sqrt{\frac{1}{3}}\Xi_c^{\prime+}\bar{D}^{(*)0}-\sqrt{\frac{2}{3}}\Sigma_{c}^{++}D_s^{(*)-}$}
					&	$\frac{1}{2}^{+}\times0^{-}$
					&	$1(\frac{1}{2})^{-}$
					&	$\frac{2}{3}g_A$
					\\
					\cline{2-2}
					\cline{4-6}

					&	$2f_{2}$
					&		
					&	\multirow{2}{*}{$\frac{1}{2}^{+}\times1^{-}$}
					&	$1(\frac{1}{2})^{-}$
					&	$-\frac{4}{45}g_A$
					\\
					\cline{2-2}
					\cline{5-6}

					&	$2f_{3}$
					&	
					&		
					&	$1(\frac{3}{2})^{-}$
					&	$\frac{13}{45}g_A$
					\\
					\hline
					%%%%%%%%%%%%%%%%%%%%%%%%%%%%%%%%%%%%%%%%%%%%%%%%%%%%%%%%%%%%%%%%%%%%%%%%%%%	
					
					\multirow{6}{*}{${P_{\psi s}^{\Sigma^{0}}}{P_{\psi s}^{\Lambda^{0}}}\pi_{0}$}
					&	$\frac{2}{\sqrt{3}}g_{1}$
					&	\multirow{3}{*}{${P_{\psi s}^{\Sigma^{0}}}:\sqrt{\frac{1}{6}}\Xi_c^{\prime+}D^{(*)-}+\sqrt{\frac{1}{6}}\Xi_{c}^{\prime0}\bar{D}^{(*)0}-\sqrt{\frac{2}{3}}\Sigma_{c}^{+}D_s^{(*)-}$}
					&	$\frac{1}{2}^{+}\times0^{-}$
					&	$1(\frac{1}{2})^{-}$
					&	$\frac{2\sqrt{3}}{15}g_A$
					\\
					\cline{2-2}
					\cline{4-6}

					&	$\frac{2}{\sqrt{3}}g_{2}$
					&		
					&	\multirow{2}{*}{$\frac{1}{2}^{+}\times1^{-}$}
					&	$1(\frac{1}{2})^{-}$
					&	$-\frac{8\sqrt{3}}{45}g_A$
					\\
					\cline{2-2}
					\cline{5-6}

					&	$\frac{2}{\sqrt{3}}g_{3}$
					&	
					&	
					&	$1(\frac{3}{2})^{-}$
					&	$-\frac{\sqrt{3}}{45}g_A$
					\\
					
					\cline{2-6}

					&	$\frac{2}{\sqrt{3}}g_{1}$
					&	\multirow{3}{*}{${P_{\psi s}^{\Lambda^{0}}}: -\sqrt{\frac{1}{2}}\Xi_c^{\prime+}D^{(*)-} + \sqrt{\frac{1}{2}}\Xi_{c}^{\prime0}\bar{D}^{(*)0} $}
					&	$\frac{1}{2}^{+}\times0^{-}$
					&	$0(\frac{1}{2})^{-}$
					&		$\frac{2\sqrt{3}}{15}g_A$
					\\
					\cline{2-2}
					\cline{4-6}

					&	$\frac{2}{\sqrt{3}}g_{2}$
					&	
					&	\multirow{2}{*}{$\frac{1}{2}^{+}\times1^{-}$}
					&	$0(\frac{1}{2})^{-}$
					&	$-\frac{8\sqrt{3}}{45}g_A$
					\\
					\cline{2-2}
					\cline{5-6}

					&	$\frac{2}{\sqrt{3}}g_{3}$
					&	
					&	
					&	$0(\frac{3}{2})^{-}$
					&	$-\frac{\sqrt{3}}{45}g_A$
					\\
					\hline
					%%%%%%%%%%%%%%%%%%%%%%%%%%%%%%%%%%%%%%%%%%%%%%%%%%%%%%%%%%%%%%%%%%%%%%%%%%%	
					\multirow{3}{*}{${P_{\psi s}^{\Sigma^{-}}}{P_{\psi s}^{\Sigma^{-}}}\pi_{0}$}
					&	$-2f_{1}$
					&	\multirow{3}{*}{${P_{\psi s}^{\Sigma^{-}}}: \sqrt{\frac{1}{3}}\Xi_c^{\prime0}D^{(*)-}-\sqrt{\frac{2}{3}}\Sigma_{c}^{0}D_s^{(*)-}$}
					&	$\frac{1}{2}^{+}\times0^{-}$
					&	$1(\frac{1}{2})^{-}$
					&	$-\frac{2}{3}g_A$
					\\
					\cline{2-2}
					\cline{4-6}

					&	$-2f_{2}$
					&	
					&	\multirow{2}{*}{$\frac{1}{2}^{+}\times1^{-}$}
					&	$1(\frac{1}{2})^{-}$
					&	$\frac{4}{45}g_A$
					\\
					\cline{2-2}
					\cline{5-6}

					&	$-2f_{3}$
					&	
					&	
					&	$1(\frac{3}{2})^{-}$
					&	$-\frac{13}{45}g_A$
					\\
					\hline
					%%%%%%%%%%%%%%%%%%%%%%%%%%%%%%%%%%%%%%%%%%%%%%%%%%%%%%%%%%%%%%%%%%%%%%%%%%%	
					
					\multirow{3}{*}{${P_{\psi ss}^{N^{0}}}{P_{\psi ss}^{N^{0}}}\pi_{0}$}
					&	$f_{1}-g_{1}$
					&	\multirow{3}{*}{${P_{\psi ss}^{N^{0}}}: \sqrt{\frac{1}{3}}\Xi_c^{\prime+}D_{s}^{(*)-}-\sqrt{\frac{2}{3}}\Omega_{c}^{0}\bar{D}^{(*)0}$}
					&	$\frac{1}{2}^{+}\times0^{-}$
					&	$\frac{1}{2}(\frac{1}{2})^{-}$
					&	$\frac{2}{15}g_A$
					\\
					\cline{2-2}
					\cline{4-6}

					&	$f_{2}-g_{2}$
					&	
					&	\multirow{2}{*}{$\frac{1}{2}^{+}\times1^{-}$}
					&	$\frac{1}{2}(\frac{1}{2})^{-}$
					&	$\frac{2}{9}g_A$
					\\
					\cline{2-2}
					\cline{5-6}

					&	$f_{3}-g_{3}$
					&	
					&	
					&	$\frac{1}{2}(\frac{3}{2})^{-}$
					&	$\frac{8}{45}g_A$
					\\
					\hline
					%%%%%%%%%%%%%%%%%%%%%%%%%%%%%%%%%%%%%%%%%%%%%%%%%%%%%%%%%%%%%%%%%%%%%%%%%%%				
					\multirow{3}{*}{${P_{\psi ss}^{N^{-}}}{P_{\psi ss}^{N^{-}}}\pi_{0}$}
					&	$g_{1}-f_{1}$
					&	\multirow{3}{*}{${P_{\psi ss}^{N^{-}}}: \sqrt{\frac{1}{3}}\Xi_c^{\prime0}D_{s}^{(*)-}-\sqrt{\frac{2}{3}}\Omega_{c}^{0}D^{(*)-}$}
					&	$\frac{1}{2}^{+}\times0^{-}$
					&	$\frac{1}{2}(\frac{1}{2})^{-}$
					&	$-\frac{2}{15}g_A$
					\\
					\cline{2-2}
					\cline{4-6}

					&	$g_{2}-f_{2}$
					&	
					&	\multirow{2}{*}{$\frac{1}{2}^{+}\times1^{-}$}
					&	$\frac{1}{2}(\frac{1}{2})^{-}$
					&	$-\frac{2}{9}g_A$
					\\
					\cline{2-2}
					\cline{5-6}

					&	$g_{3}-f_{3}$
					&	
					&	
					&	$\frac{1}{2}(\frac{3}{2})^{-}$
					&	$-\frac{8}{45}g_A$
					\\
				\toprule[1.0pt]
				\toprule[1.0pt]				
			\end{tabular}}	
		\end{table*}	
		
		\begin{table*}[htbp]
			\centering
			\caption{The axial charges of the octet hidden-charm pentaquark family in $8_{2f}$ flavor representation. The  $J_{b} ^{P_{b}}\otimes J_{m} ^{P_{m}} $ are corresponding to the angular momentum and parity of baryon and meson, respectively.}
			\label{tab_24} 
			\vspace{0.5em}
			\resizebox{180mm}{!}{	
			\begin{tabular}{ c|c|c|c|c|c}
					\toprule[1.0pt]
					\toprule[1.0pt]
				Couplings & Constants &Wave functions & $J_{b} ^{P_{b}}\otimes J_{m} ^{P_{m}} $ & $I(J^{P})$ & Results 
				\\
				\hline
				
				\multirow{3}{*}{${{}P_{\psi}^{N^{+}}}{{}P_{\psi}^{N^{+}}}\pi_{0}$}
				&	$f_{4}+g_{4}$
				&	\multirow{3}{*}{${P_{\psi}^{N^{+}}}: \Lambda_{c}^{+}\Bar{D}^{(*)0}$}
				&	$\frac{1}{2}^{+}\times0^{-}$
				&	$\frac{1}{2}(\frac{1}{2})^{-}$
				&	$0$
				\\
				\cline{2-2}
				\cline{4-6}

				&	$f_{5}+g_{5}$
				&		
				&	\multirow{2}{*}{$\frac{1}{2}^{+}\times1^{-}$}
				&	$\frac{1}{2}(\frac{1}{2})^{-}$
				&	$\frac{2}{5}g_A$
				\\
				\cline{2-2}
				\cline{5-6}

				&	$f_{6}+g_{6}$
				&
				&		
				&	$\frac{1}{2}(\frac{3}{2})^{-}$
				&	$\frac{1}{5}g_A$
				\\
				\hline
				%%%%%%%%%%%%%%%%%%%%%%%%%%%%%%%%%%%%%%%%%%%%%%%%%%%%%%%%%%%%%%%%%%%%%%%%%	
				\multirow{3}{*}{${P_{\psi}^{N^{0}}}{P_{\psi}^{N^{0}}}\pi_{0}$}
				&	$-g_{4}-f_{4}$
				&	\multirow{3}{*}{${P_{\psi}^{N^{0}}}: \Lambda_{c}^{+}D^{(*)-}$}
				&	$\frac{1}{2}^{+}\times0^{-}$
				&	$\frac{1}{2}(\frac{1}{2})^{-}$
				&	$0$
				\\
				\cline{2-2}
				\cline{4-6}

				&	$-g_{5}-f_{5}$
				&		
				&	\multirow{2}{*}{$\frac{1}{2}^{+}\times1^{-}$}
				&	$\frac{1}{2}(\frac{1}{2})^{-}$
				&	$-\frac{2}{5}g_A$
				\\
				\cline{2-2}
				\cline{5-6}

				&	$-g_{6}-f_{6}$
				&
				&	
				&	$\frac{1}{2}(\frac{3}{2})^{-}$
				&	$-\frac{1}{5}g_A$
				\\
				\hline
				%%%%%%%%%%%%%%%%%%%%%%%%%%%%%%%%%%%%%%%%%%%%%%%%%%%%%%%%%%%%%%%%%%%%%%%%%%	
				\multirow{3}{*}{${P_{\psi s}^{\Sigma^{+}}}{P_{\psi s}^{\Sigma^{+}}}\pi_{0}$}
				&	$2f_{4}$
				&	\multirow{3}{*}{${P_{\psi s}^{\Sigma^{+}}}: \Xi_{c}^{+}\Bar{D}^{(*)0}$}
				&	$\frac{1}{2}^{+}\times0^{-}$
				&	$1(\frac{1}{2})^{-}$
				&	$0$
				\\
				\cline{2-2}
				\cline{4-6}

				&	$2f_{5}$
				&		
				&	\multirow{2}{*}{$\frac{1}{2}^{+}\times1^{-}$}
				&	$1(\frac{1}{2})^{-}$
				&	$\frac{2}{5}g_A$
				\\
				\cline{2-2}
				\cline{5-6}

				&	$2f_{6}$
				&	
				&		
				&	$1(\frac{3}{2})^{-}$
				&	$\frac{1}{5}g_A$
				\\
				\hline
				%%%%%%%%%%%%%%%%%%%%%%%%%%%%%%%%%%%%%%%%%%%%%%%%%%%%%%%%%%%%%%%%%%%%%%%%%%	
				\multirow{6}{*}{${P_{\psi s}^{\Lambda^{0}}}{P_{\psi s}^{\Sigma^{0}}}\pi_{0}$}
				&	$\frac{2}{\sqrt{3}}g_{4}$
				&	\multirow{3}{*}{${P_{\psi s}^{\Sigma^{0}}}: \sqrt{\frac{1}{2}}\Xi_c^{+}D^{(*)-} + \sqrt{\frac{1}{2}}\Xi_{c}^{0}\bar{D}^{(*)0} $}
				&	$\frac{1}{2}^{+}\times0^{-}$
				&	$1(\frac{1}{2})^{-}$
				&	$0$
				\\
				\cline{2-2}
				\cline{4-6}

				&	$\frac{2}{\sqrt{3}}g_{5}$
				&		
				&	\multirow{2}{*}{$\frac{1}{2}^{+}\times1^{-}$}
				&	$1(\frac{1}{2})^{-}$
				&	$\frac{2\sqrt{3}}{15}g_A$
				\\
				\cline{2-2}
				\cline{5-6}

				&	$\frac{2}{\sqrt{3}}g_{6}$	
				&	
				&	
				&	$1(\frac{3}{2})^{-}$
				&	$\frac{\sqrt{3}}{15}g_A$
				\\
				\cline{2-6}

				&	$\frac{2}{\sqrt{3}}g_{4}$	
				&	\multirow{3}{*}{${P_{\psi s}^{\Lambda^{0}}}:\sqrt{\frac{1}{6}}\Xi_c^{+}D^{(*)-}-\sqrt{\frac{1}{6}}\Xi_{c}^{0}\bar{D}^{(*)0}-\sqrt{\frac{2}{3}}\Lambda_{c}^{+}D_s^{(*)-}$}
				&	$\frac{1}{2}^{+}\times0^{-}$
				&	$0(\frac{1}{2})^{-}$
				&	$0$
				\\
				\cline{2-2}
				\cline{4-6}

				&	$\frac{2}{\sqrt{3}}g_{5}$	
				&	
				&	\multirow{2}{*}{$\frac{1}{2}^{+}\times1^{-}$}
				&	$0(\frac{1}{2})^{-}$
				&	$\frac{2\sqrt{3}}{15}g_A$
				\\
				\cline{2-2}
				\cline{5-6}

				&	$\frac{2}{\sqrt{3}}g_{6}$	
				&	
				&	
				&	$0(\frac{3}{2})^{-}$
				&	$\frac{\sqrt{3}}{15}g_A$
				\\
				\hline
				%%%%%%%%%%%%%%%%%%%%%%%%%%%%%%%%%%%%%%%%%%%%%%%%%%%%%%%%%%%%%%%%%%%%%%%%%%	
				
				\multirow{3}{*}{${P_{\psi s}^{\Sigma^{-}}}{P_{\psi s}^{\Sigma^{-}}}\pi_{0}$}
				&	$-2f_{4}$
				&	\multirow{3}{*}{${P_{\psi s}^{\Sigma^{-}}}: \Xi_{c}^{0}D^{(*)-}$}
				&	$\frac{1}{2}^{+}\times0^{-}$
				&	$1(\frac{1}{2})^{-}$
				&	$0$
				\\
				\cline{2-2}
				\cline{4-6}

				&	$-2f_{5}$
				&	
				&	\multirow{2}{*}{$\frac{1}{2}^{+}\times1^{-}$}
				&	$1(\frac{1}{2})^{-}$
				&	$-\frac{2}{5}g_A$
				\\
				\cline{2-2}
				\cline{5-6}

				&	$-2f_{6}$
				&	
				&	
				&	$1(\frac{3}{2})^{-}$
				&	$-\frac{1}{5}g_A$
				\\
				\hline
				%%%%%%%%%%%%%%%%%%%%%%%%%%%%%%%%%%%%%%%%%%%%%%%%%%%%%%%%%%%%%%%%%%%%%%%%%%	
				
				\multirow{3}{*}{${P_{\psi ss}^{N^{0}}}{P_{\psi ss}^{N^{0}}}\pi_{0}$}
				&	$f_{4}-g_{4}$
				&	\multirow{3}{*}{${P_{\psi ss}^{N^{0}}}: \Xi_{c}^{+}D_{s}^{(*)-}$}
				&	$\frac{1}{2}^{+}\times0^{-}$
				&	$\frac{1}{2}(\frac{1}{2})^{-}$
				&	$0$
				\\
				\cline{2-2}
				\cline{4-6}

				&	$f_{5}-g_{5}$
				&	
				&	\multirow{2}{*}{$\frac{1}{2}^{+}\times1^{-}$}
				&	$\frac{1}{2}(\frac{1}{2})^{-}$
				&	$0$
				\\
				\cline{2-2}
				\cline{5-6}

				&	$f_{6}-g_{6}$
				&	
				&	
				&	$\frac{1}{2}(\frac{3}{2})^{-}$
				&	$0$
				\\
				\hline
				%%%%%%%%%%%%%%%%%%%%%%%%%%%%%%%%%%%%%%%%%%%%%%%%%%%%%%%%%%%%%%%%%%%%%%%%%%				
				\multirow{3}{*}{${{}P_{\psi ss}^{N^{-}}}{{}P_{\psi ss}^{N^{-}}}\pi_{0}$}
				&	$g_{4}-f_{4}$
				&	\multirow{3}{*}{${P_{\psi ss}^{N^{-}}}: \Xi_{c}^{0}D_{s}^{(*)-}$}
				&	$\frac{1}{2}^{+}\times0^{-}$
				&	$\frac{1}{2}(\frac{1}{2})^{-}$
				&	$0$
				\\
				\cline{2-2}
				\cline{4-6}

				&	$g_{5}-f_{5}$
				&	
				&	\multirow{2}{*}{$\frac{1}{2}^{+}\times1^{-}$}
				&	$\frac{1}{2}(\frac{1}{2})^{-}$
				&	$0$
				\\
				\cline{2-2}
				\cline{5-6}

				&	$g_{6}-f_{6}$
				&	
				&	
				&	$\frac{1}{2}(\frac{3}{2})^{-}$
				&	$0$
				\\
			\toprule[1.0pt]
			\toprule[1.0pt]
			\end{tabular}}		
		\end{table*}	
		
As presented in Table \ref{tab_23} and Table \ref{tab_24}, the axial charges of the octet hidden-charm pentaquark family are generally lower compared to that of the nucleon. In Table \ref{tab_24}, the axial charges of the pentaquark states with the spin configuration $J^{P}=\frac{1}{2}^{-}(\frac{1}{2}^{+}\otimes0^{-})$ in $8_{2f}$ flavor representation are all zero, because both $f_4$ and $g_4$ are all zero in Table \ref{tab_25}. Due to $f_5=g_5$ and $f_6=g_6$ in Table \ref{tab_25}, the axial charges of both ${P_{\psi ss}^{N^{0}}}$ and ${P_{\psi ss}^{N^{-}}}$ in $8_{2f}$ flavor representation are all zero as shown in Table \ref{tab_25}. 
		
		The axial charges at the hadron level exhibits a high sensitivity to the flavor-spin configurations of the pentaquark states. This sensitivity results in numerical discrepancies within the quark-hadron dual space, which proves advantageous for distinguishing the strong decay processes of the pentaquark states with different structures in the future. Moreover, the axial charges of the pentaquark states in Table \ref{tab_23} and Table \ref{tab_24} will be very beneficial for the theorical calculation of the pentaquark states in chiral perturbation theory.

	\section{SUMMARY}
	\label{sec5}
		With the discovery of more and more exotic hadron states comprising multiple quarks in experiments, significant progress has been made in the theoretical study of these exotic hadrons. Magnetic moments and axial charges of these exotic hadrons are intrinsic properties that offer valuable insights into their quark constituents.

		In this work, we investigate the magnetic moment of the octet hidden-charm molecular pentaquark family with quark model. Our numerical results show that the magnetic moments of the $S$-wave molecular pentaquark states in the  $ 8_{1f} $ flavor representation and $8_{2f} $ flavor representation are completely different. In the $ 8_{2f} $ flavor representation, the magnetic moments of  pentaquark states with spin configuration $J^{P}=\frac{1}{2}^{-}(\frac{1}{2}^{+}\otimes0^{-})$ are all equal to $\mu_{c}=0.38\mu_{N}$.  We notice an interesting phenomenon that the smaller the number of charges, the smaller magnetic moments of octet  hidden-charm molecular pentaquark states with the same strangeness number. The Coleman-Glashow sum-rule for the magnetic moments of pentaquark
		family turns out to hold independently of SU(3) symmetry breaking.
		
		We also calculate the axial charge of the octet hidden-charm molecular pentaquark family. Numerical results show that the axial charges of the octet hidden-charm pentaquark family are generally lower compared to that of the nucleon. The axial charges of the pentaquark states with the spin configuration $J^{P}=\frac{1}{2}^{-}(\frac{1}{2}^{+}\otimes0^{-})$ in $8_{2f}$ flavor representation are all zero. Our calculation of the pentaquark axial charges will greatly contribute to the theoretical calculation of the pentaquark states in chiral perturbation theory.
	    We hope our present study would enhance our understanding of the mechanism of hadronic molecules.
		
		\vspace{1em}
	\section*{Acknowledgments}
	This project is supported by the National Natural Science Foundation of China under Grants No. 11905171. This work is also supported by the Natural Science Basic Research Plan in Shaanxi Province of China (Grant No. 2022JQ-025) and Shaanxi Fundamental Science Research Project for Mathematics and Physics (Grant No.22JSQ016).

\end{document}